\documentclass{article}

\usepackage{geometry}
\usepackage[utf8]{inputenc}
\usepackage{authblk}
\usepackage{setspace}
\usepackage{siunitx}
\usepackage{algorithm2e}
\usepackage{amsmath, amssymb, amscd, amsthm, amsfonts, bm}
\usepackage{booktabs}
\usepackage{graphicx}
\usepackage{comment}
\usepackage{microtype}
\usepackage{hyperref}
\usepackage[numbers]{natbib}
\usepackage{xr}
\usepackage{orcidlink}
\usepackage{tabularx, threeparttable}

\DeclareMathOperator*{\argmax}{arg\,max} 

\makeatletter
\newcommand*{\addFileDependency}[1]{% argument=file name and extension
\typeout{(#1)}% latexmk will find this if $recorder=0
\@addtofilelist{#1}
\IfFileExists{#1}{}{\typeout{No file #1.}}
}\makeatother

\newcommand*{\myexternaldocument}[1]{%
\externaldocument{#1}%
\addFileDependency{#1.tex}%
\addFileDependency{#1.aux}%
}

\myexternaldocument{supplement}

\title{A generalized Bayesian stochastic block model for microbiome community detection}

\author[1*]{Kevin C. Lutz \orcidlink{0000-0002-6687-0637}}
\author[2]{Michael L. Neugent \orcidlink{0000-0002-9863-9595}}
\author[3]{Tejasv Bedi \orcidlink{0000-0001-7532-4075}}
\author[2,4]{Nicole J. De Nisco \orcidlink{0000-0002-7670-5301}}
\author[3*]{Qiwei Li \orcidlink{0000-0002-1020-3050}}

\affil[1]{Peter O'Donnell Jr. School of Public Health, The University of Texas Southwestern Medical Center, Dallas, Texas, 75390, United States}
\affil[2]{Department of Biological Sciences, The University of Texas at Dallas, Richardson, Texas, 75080, United States}
\affil[3]{Department of Mathematical Sciences, The University of Texas at Dallas, Richardson, Texas, 75080, United States}
\affil[4]{Department of Urology, The University of Texas Southwestern Medical Center, Dallas, Texas, 75390, United States}
\affil[*]{Address correspondence to: Kevin.Lutz@UTSouthwestern.edu and Qiwei.Li@UTDallas.edu}

\begin{document}

\maketitle

\begin{abstract}
Advances in next-generation sequencing technology have enabled the high-throughput profiling of metagenomes and accelerated the microbiome study. Recently, there has been a rise in quantitative studies that aim to decipher the microbiome co-occurrence network and its underlying community structure based on metagenomic sequence data. Uncovering the complex microbiome community structure is essential to understanding the role of the microbiome in disease progression and susceptibility. Taxonomic abundance data generated from metagenomic sequencing technologies are high-dimensional and compositional, suffering from uneven sampling depth, over-dispersion, and zero-inflation. These characteristics often challenge the reliability of the current methods for microbiome community detection. To this end, we propose a Bayesian stochastic block model to study the microbiome co-occurrence network based on the recently developed modified centered-log ratio transformation tailored for microbiome data analysis. Our model allows us to incorporate taxonomic tree information using a Markov random field prior. The model parameters are jointly inferred by using Markov chain Monte Carlo sampling techniques. Our simulation study showed that the proposed approach performs better than competing methods even when taxonomic tree information is non-informative. We applied our approach to a real urinary microbiome dataset from postmenopausal women, the first time the urinary microbiome co-occurrence network structure has been studied. In summary, this statistical methodology provides a new tool for facilitating advanced microbiome studies.

%\keywords{Bayesian inference, stochastic block model, Markov random field, microbiome co-occurrence network, community detection, taxonomic tree}
\end{abstract}

\section{Introduction}

The term \textit{microbiome} was first introduced by Nobel Laureate Joshua Lederberg \citep{lederberg2001ome, liu2016focus} and refers to the collective genomes of microorganisms or the microorganisms themselves \citep{turnbaugh2007human}. Ecological interactions of these microorganisms are important because they affect microbiome function and host health through the formation of complex microbiome communities \citep{antwis2017fifty}. Uncovering these relationships is essential to understanding the role of the microbiome in disease progression and susceptibility \citep{hall2019co, shreiner2015gut}. For example, microbial interactions in the human gut microbiome have been associated with the progression of several diseases such as colorectal cancer \citep{marchesi2011towards}, diabetes \citep{karlsson2013gut}, and inflammatory bowel disease \citep{halfvarson2017dynamics}. \citet{hall2019co} found evidence that the network of microbes in the human gut microbiome is composed of distinct communities that co-occur and interact with one another. Further, it was found that each community tends to have similar metagenomic functional properties. Thus, uncovering the underlying community structure of a microbiome network is the key to understanding its impact on human health \citep{human2012structure}. From this point forward, a network of microbial interactions will be referred to as a \textit{microbiome co-occurrence network}.

%Studying the structure of the microbiome is already an active area of microbiome research. 
\textit{Network analysis} is a widely used statistical method that infers the complex structure and associations among entities such as persons or microbes \citep{hevey2018network}. A graphical representation of a network consists of nodes and edges. In microbiome research, each node is a taxon and an existing edge represents a significant association between two taxa. An inferred microbiome co-occurrence network can help characterize taxon-taxon associations and reveal their latent properties, mechanisms, and structures \citep{jiang2020harmonies}. Co-occurrence research within microbiomes usually considers taxon-taxon associations, which means all microbes are from the same level of the taxonomic tree hierarchy \citep{lutz2022survey}. Network analysis methods use either a similarity metric or a model-based approach to determine these associations. For further information, \citet{lutz2022survey} provide a detailed survey of available methods for microbiome network analysis. 

\textit{Community detection} is one of the fundamental problems in network analysis. The most widely used model-based approach for performing community detection is the \textit{stochastic block model} (SBM) \citep{karrer2011stochastic}, which was first introduced by \citet{holland1983stochastic}. SBM has been employed in a broad range of applications including medicine \citep{li2021bayesian}, social media \citep{yu2018detecting}, sociology \citep{mcdaid2013improved}, political science \citep{latouche2011overlapping}, military strategy \citep{olivella2022dynamic}, infrastructure \citep{yu2020modeling}, and many more. Applications in microbiome co-occurrence networks of bacteria, genes, or proteins include microbial communities associated with pH in arctic soil \citep{faust2016conet}, taxon-taxon communities of the human gut microbiome associated with disease development \citep{marchesi2016gut,hall2019co}, protein-protein interactions associated with pancreatic cancer \citep{stanley2019stochastic}, optimizing treatment plant operations by understanding microbial communities in wastewater \citep{de2022niche}, controlling tick-related diseases by understanding the interactions of tick-borne microbial communities over time \citep{lejal2021temporal}, and the use of protein-protein communities related to SARS-CoV-2 to better understand COVID-19 \citep{ghavasieh2021multiscale}. Applications using frequentist \citep{latouche2011overlapping,abbe2015exact,yu2018detecting,stanley2019stochastic} and Bayesian \citep{mcdaid2013improved,faskowitz2018weighted,hall2019co,nouedoui2013bayesian} approaches are available in the literature. Recently, SBMs have been used to uncover the underlying microbiome community structure by clustering all taxa in the microbiome co-occurrence network based on their connectivity patterns \citep{hall2019co, cao2021effects, bhar2022application}. In general, SBMs aim to partition the nodes (e.g., taxa) of a network with heterogeneous connectivity patterns into mutually exclusive blocks (e.g., communities) with homogeneous connectivity patterns \citep{mcdaid2013improved, aicher2015learning, lee2019review}. Furthermore, SBMs can infer the latent underlying structural patterns of a network \citep{lee2019review} and estimate the edge probabilities within each block and between blocks \citep{aicher2015learning}.
%$K\times K$ matrix containing the probabilities of edges connected to nodes within and between homogeneous blocks. 
 %Generally, the frequentist approach maximizes the data likelihood; whereas, the Bayesian approach maximizes the posterior distribution \citep{hoff2009first}. 
Table \ref{tab:sbm_software} in the appendix provides a non-exhaustive catalog of available SBM software package information for users in \texttt{R}, \texttt{Python}, and \texttt{C++}, while Table \ref{tab:sbm_url} lists the websites for software access and documentation. 

%The Bayesian approach is more flexible because prior information can be incorporated into the model, has no restrictive assumptions, relies on probability distributions that can handle high-dimensional data, does not need to assume a fixed number of communities, and determines community membership using posterior probability \citep{nowicki2001estimation, hoff2009first}. 

We searched the literature for SBMs applied specifically to microbiome co-occurrence network analysis \citep{hall2019co, bhar2022application, aguinaldo2021namdata, dunphy2019structure, kuntal2019netshift,stanley2019stochastic, connor2017using, widder2022metagenomic,wahid2022literature}, and found that all taxa in the network were from only one level of the taxonomic tree hierarchy (e.g., species). Thus, none of them account for taxonomic tree structure. Integrating the available taxonomic or phylogenetic tree structure into SBMs could provide additional information about microbiome communities because taxa (e.g., species) having the same parent (e.g., genus) or with similar functional properties tend to cluster together \citep{hall2019co,won2020can,ozen2012defining,xu2021genome}. In addition, most of those SBMs ignore multiple characteristics of taxonomic abundance data generated from metagenomic sequencing technology \citep{cullen2020emerging}, such as high-dimensionality, zero-inflation, over-dispersion, and compositionality \citep{badri2018normalization,jiang2021bayesian}, resulting in information loss and inference bias \citep{quinn2019field}. Thus, tailored statistical methods accounting for those challenging characteristics are required. 

%For example, \citet{dunphy2019structure} normalized their data using DeSeq2 \citep{love2014moderated} and then used linear methods (e.g., Pearson correlation) to estimate the network prior to performing community detection. DeSeq2 normalization does not account for zero inflation and Pearson correlation does not account for non-linearity in the transformed data. 
%The common denominator of these applications is that each microbiome co-occurrence network contains features from only one level of the taxonomic tree (e.g., species only). 

%The above issues motivated us to update the standard Bayesian SBM for performing community detection on a binary network to account for available taxonomic tree information while properly addressing the challenges posed by the abundance data. As a result, 
This paper proposes a generalized Bayesian SBM with a Markov random field (MRF) prior, which we refer to as Bayesian-SBM-MRF. The MRF prior accounts for taxonomic tree structure by allowing the model to incorporate more than one level of the taxonomic tree, which is an attractive and novel feature of our proposed model. Our proposed model considers two sources of binary information to perform microbiome community detection: (i) the taxon-taxon microbiome co-occurrence network and (ii) taxonomic tree information. Unlike other SBMs in the literature for microbiome study, our model applies the recently developed modified centered-log ratio (MCLR) transformation \citep{yoon2019microbial} to account for zero-inflation and compositionality in the taxonomic abundance data. Additionally, we use the Spearman correlation coefficient of the MCLR-transformed abundances to identify non-linear pairwise taxon-taxon associations. We show in our simulations that Bayesian-SBM-MRF performs at least as good if not better when we incorporate both (i) and (ii) instead of just only (i) in the analysis, and significantly outperforms other existing SBMs. We implement our model on a real urinary microbiome dataset taken from a controlled, cross-sectional study of recurrent urinary tract infections (rUTI) from post-menopausal women. Additionally, this is the first time that the urinary microbiome community structure has been investigated. We also implemented our model on a second real dataset based on the characters in the classic novel \textit{Les Mis\'erables} to demonstrate its broad use. Our model provides a new tool for advanced studies of microbiome co-occurrence networks.
%Further, {Bayesian-SBM-MRF} outperforms other existing SBMs since they only take (i) into account. Notably, our model also shows superior performance even when (ii) is non-informative. 

%The novelty of our method is two-fold. First, \citet{yoon2019microbial} proposed MCLR for the purpose of estimating a network (more information is in Section \ref{sbm_data}); whereas, we apply MCLR for the purpose of performing community detection on a microbiome co-occurrence network. As far as we know, no one else in the literature has used the MCLR transformation to perform community detection on a microbiome co-occurrence network. Second, we account for available taxonomic tree information through the use of the MRF prior to incorporate more than one taxonomic level to perform community detection. 

The remainder of the article is organized as follows: Sections \ref{sbm_data} and \ref{joint_data_model} introduce the data preprocessing, the standard Bayesian SBM, and the proposed generalized SBM with an MRF prior; Section \ref{joint_model_fitting} describes the Markov chain Monte Carlo (MCMC) algorithms for model fitting and Bayesian posterior inference; Section \ref{results} provides the results of a simulation study to assess and compare the proposed model to current methods as well as results from the analysis on two real datasets; Section \ref{conclusion} concludes the article with a summary and discussion of the proposed {Bayesian-SBM-MRF} model.

\section{Data Preparation}\label{sbm_data}
In this section, we describe the two sources of binary taxonomic information to perform microbiome community detection; namely, the microbiome co-occurrence network and taxonomic tree, as illustrated in Figure \ref{fig:sbm_graphic}.

Let $\boldsymbol{Y}=[y_{ij}]\in\mathbb{N}^{n\times p}$ denote an $n\times p$ taxonomic abundance matrix, with $y_{ij}$ indicating the count of taxon $j$ observed from sample $i$, $i=1,\ldots,n,j=1,\ldots,p$. We use $\boldsymbol{y}_{i\cdot}=(y_{i1},\ldots,y_{ip})^\top$ and $\boldsymbol{y}_{\cdot j}=(y_{1j},\ldots,y_{nj})^\top$ to denote the vector from the $i^{\text{th}}$ row and $j^{\text{th}}$ column of $\boldsymbol{Y}$, respectively. We use the same formatting for any matrix throughout this paper. 

\subsection{Microbiome co-occurrence network}
Let $\boldsymbol{X}=[x_{ij}]\in[0,1]^{n\times p}$ denote the $n\times p$ matrix of the relative abundances (i.e., compositions), where $x_{ij}=y_{ij}/\sum_{j=1}^py_{ij}$. The vector of relative abundances in the $i^{\text{th}}$ sample, $\boldsymbol{x}_{i\cdot}$, is defined on a $p$-dimensional simplex (i.e., $x_{ij}\ge0,\forall j$ and $\sum_{j=1}^px_{ij}=1$). To map a composition to a Euclidean vector space, \citet{aitchison1982statistical} proposed the CLR transformation, which is to scale the relative abundances $\boldsymbol{x}_{i\cdot}$ by its geometric mean and then take the logarithm to remove the unit-sum constraint. However, taxonomic abundance data contain a large proportion of zero counts attributed to rare or low-abundance taxa that may be present in only a small percentage of samples; whereas, others are not recorded due to the limitations of the sampling effort. The CLR transformation adds an arbitrary pseudo value to both non-zero and zero values, which disguises the zeros and may lead to spurious correlations between taxa because zeros and non-zeros are treated equally. To remedy this issue, \citet{yoon2019microbial} recently proposed MCLR, which transforms only the non-zero values, and they demonstrated that MCLR reduces bias and variance compared to CLR. Specifically, the MCLR transformation, given by Equation \eqref{mclr}, takes the log of the ratio of each $x_{ij}\neq 0$ and the geometric mean of all non-zeros in $\boldsymbol{x}_{i\cdot}$. Let $\boldsymbol{V}=[v_{ij}]\in\mathbb{R}^{n\times p}$ denote the $n\times p$ matrix of the MCLR-transformed relative abundances, each element of which is expressed as
\begin{equation}\label{mclr}
    v_{ij}=\begin{cases}
        \begin{array}{lc}
            0 & \text{ if } x_{ij}=0 \\
            \log(x_{ij}/\Tilde{g}(\boldsymbol{x}_{i\cdot}))+\epsilon_i & \text{ if } x_{ij}\neq0
        \end{array}
    \end{cases},
\end{equation}
where $\Tilde{g}(\boldsymbol{x}_{i\cdot})=\left(\prod_{j=1}^p x_{ij}^{I(x_{ij}\neq0)} \right)^{\frac{1}{\sum_{j=1}^p I(x_{ij}\neq0)}}$ gives the geometric mean of the non-zero relative abundances in sample $i$, where $I(\cdot)$ is the indicator function. MCLR reduces to robust CLR \citep{martino2019novel} when $\epsilon_i=0,\forall i$. Alternatively, we can make all non-zero entries strictly positive by letting $\epsilon_i=1+|\text{min}_{\{j:x_{ij}\neq0\}}\{\log(x_{i1}/\Tilde{g}(\boldsymbol{x}_{i\cdot})),\ldots,\log(x_{ip}/\Tilde{g}(\boldsymbol{x}_{i\cdot}))\}|$ as suggested by \citet{yoon2019microbial}, which imposes a shift above zero on the transformed values. 

While the log transformation is a feature of MCLR, it does not guarantee linearization \citep{ng2019gamma}, especially since the transformation only applies to each $x_{ij}\neq 0$. The data remain positively skewed after transformation due to the non-transformed zeros. As a result, to measure the similarity between any pair of taxa $j$ and $j'$ in terms of MCLR-transformed relative abundances (i.e., between $\boldsymbol{v}_{\cdot j}$ and $\boldsymbol{v}_{\cdot j'}$), we use their Spearman correlation coefficient $\rho_{jj'}$ to account for non-linearity \citep{corder2014nonparametric}. We further test the null hypothesis $\rho_{jj'}=0$ versus the alternative $\rho_{jj'}\neq0$ and obtain a $p$-value to determine the significance. All $p$-values are further adjusted using the Benjamini-Hochberg procedure to control the false discovery rate \citep{benjamini1995controlling}. To construct the microbiome co-occurrence network represented by a $p\times p$ binary adjacency matrix $\boldsymbol{G}=[g_{jj'}]\in\{0,1\}^{p\times p}$, we assigned an edge as $g_{jj'}=1$ if the corresponding adjusted $p$-value of $\rho_{jj'}$ is less than the significance level of $\alpha=0.05$ and zero otherwise. 

\subsection{Taxonomic tree}
We make use of taxonomic tree information by indicating if two taxa (e.g., species) have the same parent (e.g., genus). A tree is an undirected graph where any two nodes are connected by exactly one path. Thus, we describe the taxonomic tree using
an adjacency matrix. Let $\boldsymbol{Q}=[q_{jj'}]\in\{0,1\}^{p\times p}$ denote a $p\times p$ binary adjacency matrix where $q_{jj'}=1$ indicates that taxa $j$ and $j'$ have the same parent and zero otherwise for $j\neq j'$. The information from $\boldsymbol{Q}$ allows the model to include the taxonomic tree information, which is an attractive feature as well as one of the novelties of our model. As a caution, we recommend incorporating only genus or family level at most as parents. If you incorporate information too far up the taxonomic tree, then all the taxa will naturally collapse into one group, rendering useless results. While both $\boldsymbol{G}$ and $\boldsymbol{Q}$ are binary adjacency matrices, experts agree that very little information is lost when sequence data are converted to the binary level \citep{tuna2009classification}. For example, binary data have been used for the analysis of gene expression data and have produced reasonable results \citep{tuna2009classification,giovanini2022stochastic}. 

%A $p \times p$ binary undirected network, denoted as $\boldsymbol{G},$ is then estimated from the taxonomic abundance data $\boldsymbol{Y}$. For simplicity, the adjacency matrix is also denoted as $\boldsymbol{G}$ where $\boldsymbol{G}=[g_{jj'}]\in\{0,1\}^{p\times p}$ is a $p\times p$ binary matrix. Let $g_{jj'}=1$ indicate the existence of an edge between taxa $j$ and $j'$ and $g_{jj'}=0$ otherwise for $j\neq j'$. To estimate $\boldsymbol{G},$ we first applied the MCLR transformation to the relative abundances and then calculated the Spearman correlation coefficient of each transformed taxon-taxon pair.  

\begin{figure}[!h]
    \centering
    \includegraphics[width=1.0\textwidth]{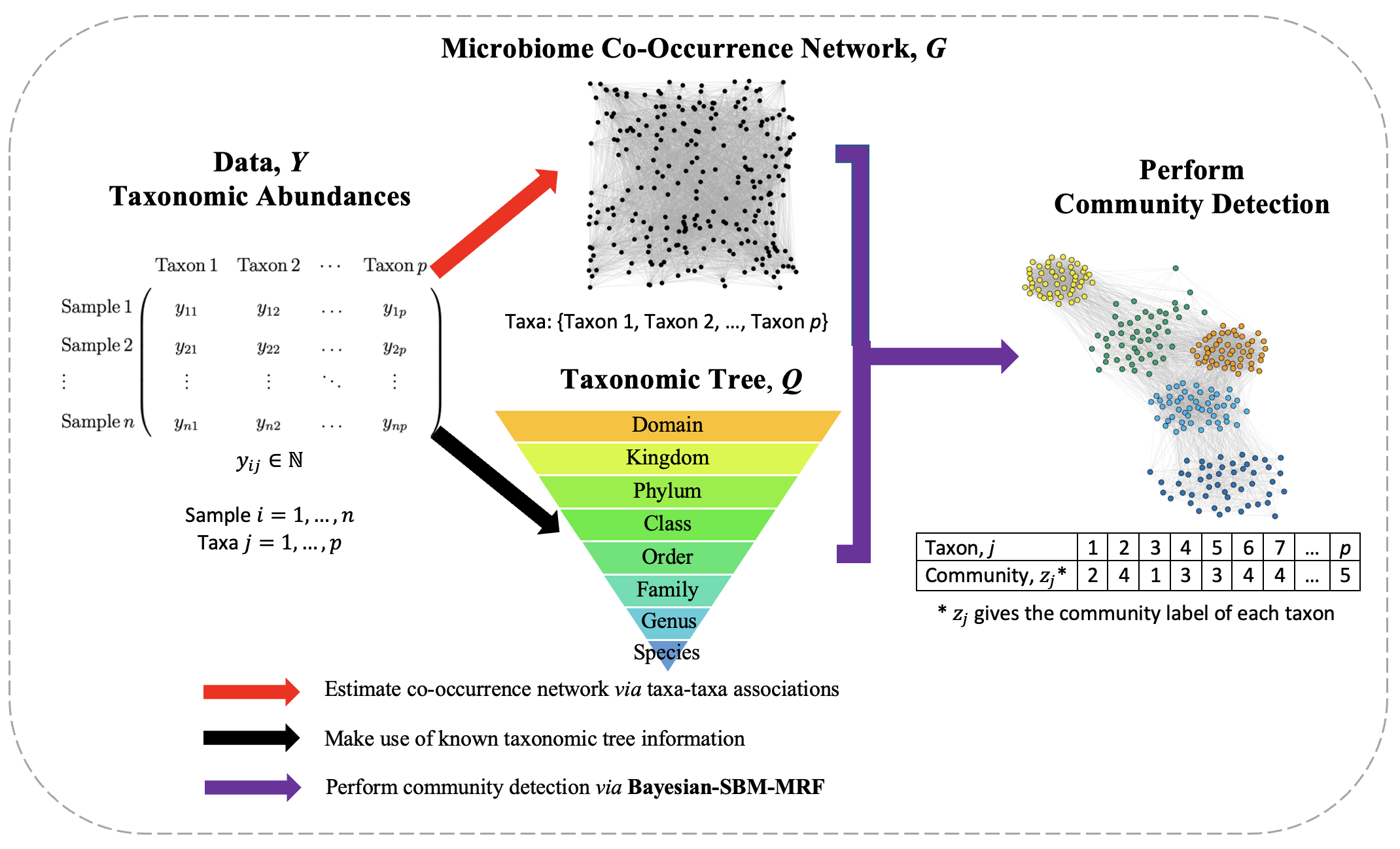}
    \caption{An illustration of {Bayesian-SBM-MRF} workflow. First, the microbiome co-occurrence network $\boldsymbol{G}$ is estimated from the taxonomic abundance data with $n$ samples and $p$ taxa using a pairwise similarity metric (red arrow). The taxonomic tree $\boldsymbol{Q}$ paired with the taxonomic abundance data is extracted (black arrow). Second, community detection is performed using {Bayesian-SBM-MRF} to infer the underlying microbiome community structure, indicated by $\boldsymbol z,$ by integrating $\boldsymbol{G}$ and $\boldsymbol{Q}$ (purple arrow).}
    \label{fig:sbm_graphic}
\end{figure}
 
\section{Model}\label{joint_data_model}
In this section, we first review the standard Bayesian SBM model in Section \ref{sbm_model} and then provide the full details about our proposed {Bayesian-SBM-MRF} in Section \ref{sbm_mrf}. 

\subsection{A Review of the Bayesian Stochastic Block Model}\label{sbm_model}
The standard Bayesian SBM for a binary network \citep{mcdaid2013improved,nowicki2001estimation} performs community detection on $p$ taxa to cluster them into $K$ homogeneous communities \textit{via} a finite Bernoulli mixture model that depends on the community indicator vector $\boldsymbol{z}$ and the edge probability matrix $\boldsymbol{\Omega}$, which are independent parameters by assumption. Note that the number of communities is assumed to be fixed and cannot exceed $n$. Let $\boldsymbol{z}=[z_j]\in\{1,\ldots,K\}^{p\times 1}$ denote the community labels for taxon $1,\ldots,p$, where $z_j=k$ indicates that taxon $j$ belongs to community $k$. The total number of taxa belonging to community $k$ is denoted by $n_k=\sum_{j=1}^p I(z_j=k)$. Next, $\boldsymbol{\Omega}=[\omega_{kk'}]\in [0,1]^{K\times K}$ contains the edge probabilities, where each diagonal element $\omega_{kk}$ and each off-diagonal element $\omega_{kk'},k\neq k'$ indicate the probability of observing an edge between any two taxa within community $k$ and between communities $k$ and $k'$, respectively.  

We assume that the presence of an edge between taxa $j$ and $j'$ in the microbiome co-occurrence network is a Bernoulli random variable conditional on their community memberships,
\begin{equation}
    g_{jj'}|z_j=k,z_{j'}=k',\omega_{kk'}\sim\text{Bern}(\omega_{kk'}).
\end{equation}
Thus, we can express the full data-likelihood as
\begin{align}\label{smb_loglike}
    \begin{split}
            f(\boldsymbol{G}|\boldsymbol{z},\boldsymbol{\Omega}) &=\prod_{k=1}^K\prod_{\{j<j':z_j=k,z_{j'}=k\}}\text{Bern}(g_{jj'}|\omega_{kk})  \prod_{k<k'}\prod_{\{j<j':z_j=k,z_{j'}=k'\}}\text{Bern}(g_{jj'}|\omega_{kk'}) \\
            %&= \prod_{k=1}^K\prod_{\{j<j':z_j=k,z_{j'}=k\}} \omega_{kk}^{g_{jj'}}(1-\omega_{kk})^{1-g_{jj'}}  \prod_{k<k'}\prod_{\{j<j':z_j=k,z_{j'}=k'\}} \omega_{kk'}^{g_{jj'}}(1-\omega_{kk'})^{1-g_{jj'}}  \\
            &= \prod_{k=1}^K \omega_{kk}^{M_{kk}}(1-\omega_{kk})^{N_{kk}-M_{kk}} \prod_{k<k'}\omega_{kk'}^{M_{kk'}}(1-\omega_{kk'})^{N_{kk'}-M_{kk'}},
    \end{split}
\end{align}
where $M_{kk'}$ and $N_{kk'}$ are the number of observed and total possible edges, respectively, within community $k$ or between communities $k$ and $k'$. Specifically,
\begin{equation}\label{N}
N_{kk'}=\begin{cases}
    \begin{array}{ll}
            {n_k\choose 2} & \text{ if } k = k'  \\
             n_kn_{k'} & \text{ if } k\neq k'
        \end{array}
\end{cases} 
\text{   and   }
    M_{kk'}=\sum_{j=1}^p I(g_{jj'}=1) I(z_j=k)I(z_{j'}=k').
\end{equation}

We model the latent community membership for each taxon, $z_j$, from a multinomial distribution, which is expressed as
\begin{equation}\label{sbm_prior}
    z_j|\boldsymbol\pi \sim \text{Mult}(1,\boldsymbol\pi),
\end{equation}
where $\boldsymbol\pi=(\pi_{1},\ldots,\pi_{K})^\top$ with $\pi_k$ indicating the size of community $k$ \textit{a priori}. Next, $\boldsymbol{\pi}$ is assumed to be a random variable.  So, we place a Dirichlet (Dir) prior on $\boldsymbol\pi$, i.e., $\boldsymbol\pi\sim\text{Dir}(\boldsymbol\alpha),$ where $\boldsymbol\alpha=(\alpha_{1},\ldots,\alpha_{K})^\top$ is a positive real-valued vector. Without any specific reason, each $\alpha_{k}$ is usually set to $1$ to obtain a uniform hyperprior \citep{hoff2009first}. %In other words, $\pi(\boldsymbol{\pi}|\boldsymbol{\alpha})=1.$ The parameters $\boldsymbol\Omega \text{ and } \boldsymbol{z}$ are assumed to be independent. 
Thus, the full conditional posterior density for $z_j$ belonging to community $k$ given everything else can be expressed as
\begin{align}\label{post_z}
    \begin{split}
        \pi(z_j = k | \boldsymbol{z}_{-j},\boldsymbol\Omega,\boldsymbol{G})
        &\propto f(\boldsymbol{G}|\boldsymbol{z},\boldsymbol\Omega) \pi(z_j=k,\boldsymbol{z}_{-j}|\boldsymbol\pi)\pi(\boldsymbol\pi)\\
        &= \prod_{k=1}^K \omega_{kk}^{M_{kk}}(1-\omega_{kk})^{N_{kk}-M_{kk}} \prod_{k<k'}\omega_{kk'}^{M_{kk'}}(1-\omega_{kk'})^{N_{kk'}-M_{kk'}}\\
        &\,\,\,\,\,\,\,\,\,\times\prod_{k=1}^K\frac{n!}{n_1!\cdots n_K!}\pi_k^{n_k},
    \end{split}
\end{align}
where $\boldsymbol{z}_{-j}$ denotes all the elements in $\boldsymbol{z}$ excluding the $j^{\text{th}}$ element. Let $\boldsymbol{\eta}=(\eta_1,\ldots,\eta_K)^\top$ denote the normalized posterior probability vector for community membership, where $\eta_k={ \pi(z_j = k | \boldsymbol{z}_{-j},\boldsymbol\Omega,\boldsymbol{G})}/{\sum_{m=1}^K  \pi(z_j = m | \boldsymbol{z}_{-j},\boldsymbol\Omega,\boldsymbol{G})}$. Then, we can sample each community membership label, $z_j$, from a multinomial distribution
\begin{equation}\label{classes_sbm}
    z_j|\boldsymbol{z}_{-j},\boldsymbol\Omega,\boldsymbol{G}\sim\text{Mult}(1, \boldsymbol\eta).
\end{equation}
%using the posterior probabilities calculated by Equation \eqref{normprob}.
To complete the model specification, we impose a beta prior on each $\omega_{kk'}$, i.e., $\omega_{kk'}\sim\text{Beta}(a_{\omega},b_{\omega})$, where $a_{\omega}\text{ and }b_{\omega}$ are fixed hyperparameters. A common non-informative setting is $a_{\omega}=b_{\omega}=1$ \citep{hoff2009first}. This conjugate setting results in a beta distribution for the posterior distribution, $\omega_{kk'}|\boldsymbol{z},\boldsymbol{G} \sim\text{Beta}(M_{kk'}+a_{\omega},N_{kk'}-M_{kk'}+b_{\omega})$. 

\subsection{Generalized Bayesian Stochastic Block Model with a Markov Random Field Prior}\label{sbm_mrf}
Here, we propose {Bayesian-SBM-MRF}: the generalized version of the standard Bayesian SBM model from Section \ref{sbm_model}. {Bayesian-SBM-MRF} integrate two different types of taxonomic information, $\boldsymbol{G}$ and $\boldsymbol{Q}$, illustrated in Figure \ref{fig:sbm_graphic}. 
%The first step (red arrow) uses the taxonomic abundance data $\boldsymbol{Y}$ to estimate the microbiome co-occurrence network $\boldsymbol{G}$, which was described in Section \ref{sbm_data}. The taxonomic tree information $\boldsymbol{Q}$ (black arrow) is not an additional step but rather is additional information about the data. The second step (purple arrow) performs community detection using our proposed {Bayesian-SBM-MRF} by incorporating two sources of information: the taxon-taxon co-occurrence network $\boldsymbol{G}$ and the taxonomic tree information $\boldsymbol{Q}$. 
We propose to replace the multinomial prior shown in Equation \eqref{sbm_prior} with an MRF prior, which can incorporate information from the given taxonomic tree on microbiome community detection.
%The Ising model, named after Ernst Ising, was the prototype to the MRF \citep{besag1986statistical} and focused on the physics of phase transitions \citep{cipra1987introduction}. For example, a phase transition in physical science occurs when water changes from liquid to solid when the temperature goes below $32^{\circ}F.$  A small change in temperature changes the state of the entire matter. So, a phase transition in a statistical model is concerned with how a small change in a model hyperparameter can change the state or quality of overall model performance. 
The MRF \citep{besag1974spatial} is a class of parametric models used for spatial data analysis that originated in theoretical physics \citep{clifford1990markov, cipra1987introduction}. %The term \textit{spatial} has broad meaning. Examples of spatial information include coordinates in the $xy$-plane or the level of hierarchy in the taxonomic tree where there are no coordinates. 
Essentially, the MRF is a nearest neighbor problem where we are interested in calculating the conditional probability that a particular taxon $j$ belongs to community $k$ (i.e., $z_j=k$) given all neighboring taxa. In this paper, a neighbor is defined as any two taxa that have the same parent such as genus. Thus, the MRF prior offers a way to incorporate the taxonomic tree information $\boldsymbol{Q}$ by encouraging two taxa with the same parent to be clustered in the same community. %The MRF prior is of the exponential family \citep{kendall1961advanced}, fully detailed in \citet{besag1974spatial} and \citet{besag1986statistical}. 
In particular, we write the conditional probability density for $z_j$ belonging to community $k$ given all other taxa memberships $\boldsymbol{z}_{-j}$ as
\begin{align}\label{MRFprior}
    \begin{split}
        z_j|\boldsymbol{z}_{-j},\boldsymbol{Q}&\sim\text{MRF}(e_k,f),\\
        \pi(z_j=k|\boldsymbol{z}_{-j},\boldsymbol{Q})&\propto \exp\left(e_k+f\sum_{\{j': q_{jj'}=1\}}I(z_{j'}=k) \right),
    \end{split}
\end{align}
where $f\in\mathbb{R}_{\geq 0}$ and $e_k=\log(\eta_k)$. The MRF prior in Equation \eqref{MRFprior} reduces to the prior in Equation \eqref{sbm_prior} when $f=0,$ which means that the standard Bayesian SBM in Section \ref{sbm_model} is a special case of the proposed generalized model. Additionally, the prior reduces to a non-informative discrete uniform prior when $f=0$ and $e_k=\log(\eta_k)=\log(1/K)$. For that reason, we set $e_k=\log(1/K)$ in our model. When $f>0$, the available taxonomic tree information is incorporated into the model from $\boldsymbol{Q}$. When $f$ is too large, then $\pi(z_j=k|\boldsymbol{z}_{-j})\to\infty$ and the model undergoes phase transition. A phase transition in a statistical model is concerned with how a small change in a model hyperparameter (e.g., $f$) can change the state or quality of overall model performance. In our simulation in Section \ref{sbm_simulation2}, we determined that $f=1$ is a reasonable value for the MRF prior setting. Then, the full conditional posterior density for the community labels from Equation \eqref{post_z} is updated as 
        \begin{align}\label{mrf_density}
    \begin{split}
       \pi(z_j=k|\boldsymbol{z}_{-j},\boldsymbol{\Omega},\boldsymbol{G},\boldsymbol{Q})&\propto f(\boldsymbol{G}|\boldsymbol{z},\boldsymbol\Omega)\pi(z_j=k|\boldsymbol{z}_{-j},\boldsymbol{Q}) \\
       &\propto\prod_{k=1}^K \omega_{kk}^{M_{kk}}(1-\omega_{kk})^{N_{kk}-M_{kk}} \prod_{k<k'}\omega_{kk'}^{M_{kk'}}(1-\omega_{kk'})^{N_{kk'}-M_{kk'}} \\
       &\,\,\,\,\,\,\,\times\exp\left(f\sum_{\{j': q_{jj'}=1\}}I(z_{j'}=k) \right)
    \end{split}.
\end{align}
Let $\boldsymbol{\xi}=(\xi_1,\ldots,\xi_K)^\top$ denote the normalized posterior probability vector for community membership, where $\xi_k={ \pi(z_j = k | \boldsymbol{z}_{-j},\boldsymbol\Omega,\boldsymbol{G},\boldsymbol{Q})}/{\sum_{m=1}^K  \pi(z_j = m | \boldsymbol{z}_{-j},\boldsymbol\Omega,\boldsymbol{G},\boldsymbol{Q})}$. Then, we can sample each community membership label, $z_j$, from a multinomial distribution
\begin{equation}\label{classes_sbm2}
    z_j|\boldsymbol{z}_{-j},\boldsymbol\Omega,\boldsymbol{G},\boldsymbol{Q}\sim\text{Mult}(1, \boldsymbol\xi).
\end{equation}
We follow the standard Bayesian SBM to update the edge probabilities, $\omega_{kk'}|\boldsymbol{z},\boldsymbol{G} \sim\text{Beta}(M_{kk'}+a_{\omega},N_{kk'}-M_{kk'}+b_{\omega})$, because they are conditionally independent of the community label $\boldsymbol{z}$ and irrelevant to the taxonomic tree $\boldsymbol{Q}$.

\section{Model Fitting}\label{joint_model_fitting}
In Section \ref{joint_mcmc}, we first describe the details of the MCMC algorithms based on a two-step Gibbs sampler \citep{geman1984stochastic}. % to fit the model and sample the model parameters, $\boldsymbol\Omega$ and $\boldsymbol{z}$. 
Then, we give the details of posterior inference for the parameters of main interest, $\boldsymbol\Omega$ and $\boldsymbol{z}$, as well as how to select the number of communities $K$ in Section \ref{inference}.

\subsection{MCMC Algorithms}\label{joint_mcmc}
Posterior sampling can be easily implemented for {Bayesian-SBM-MRF} using a two-step Gibbs sampler since both $\boldsymbol{z}$ and $\boldsymbol\Omega$ can be sampled from multinomial and beta distributions, respectively. The Gibbs sampler will be run for $T$ iterations where we discard the first half of posterior samples as burn-in samples. The number of burn-in samples is $B=T/2$ for iterations $t=1,\ldots,B$. Then, the total number of after burn-in posterior samples is $T-B$ for iterations $t=B+1,\ldots,T$. Algorithm \ref{algo:algorithm} illustrates the model fitting steps for  {Bayesian-SBM-MRF} where parameters $\boldsymbol{z}$ and $\boldsymbol{\Omega}$ are jointly inferred using a Gibbs sampler.
\SetKwComment{Comment}{/* }{ */}
\begin{algorithm}
\caption{Gibbs Sampler for {Bayesian-SBM-MRF}}
\label{algo:algorithm}
\KwData{$\boldsymbol{G},\boldsymbol{Q}$; fix $a_\omega,b_\omega, K, T, f$}
\textbf{Initialize: }$\boldsymbol{z}$\\
\For{$t$ in $1:T$}{
   \For{$k$ in $1:K$}{
   \If{$k\leq k'$}{
    $\omega_{kk'}^{(t+1)}|\boldsymbol{z}^{(t)},\boldsymbol G \sim \text{Beta}(a_\omega + M_{kk'}^{(t)}, b_\omega + N_{kk'}^{(t)}-M_{kk'}^{(t)})$\Comment*[r]{Update $\boldsymbol\Omega$}
    }
   }
   \For{$j$ in $1:p$}{
      \For{$k$ in $1:K$}{
        %$z_j^{(t+1)}=k|\boldsymbol{z}_{-j}^{(t)}\sim\text{MRF}(e_k,f)$
         
         $ \xi_{k} = {\pi(z_j^{(t+1)}=k|\boldsymbol{z}_{-j}^{(t+1)},\boldsymbol{\Omega},\boldsymbol{G},\boldsymbol{Q})}/{\sum_{m=1}^K\pi(z_j^{(t+1)}=m|\boldsymbol{z}_{-j}^{(t)},\boldsymbol{\Omega},\boldsymbol{G},\boldsymbol{Q})}$}
          $z_j^{(t+1)}|\boldsymbol{z}_{-j}^{(t+1)},\boldsymbol{\Omega},\boldsymbol{G},\boldsymbol{Q})\sim \text{Mult}(1; \boldsymbol\xi)$  \Comment*[r]{Update $\boldsymbol{z}$} 
         }
          \textbf{Store: } $\boldsymbol{\Omega}^{(t+1)}, \boldsymbol{z}^{(t+1)}$;
      }
\end{algorithm}

\subsection{Posterior Inference}\label{inference}
We make three inferences that will provide a comprehensive scope of the community structure of the microbiome co-occurrence network from the resulting posterior samples. The first is to infer the edge probabilities in $\boldsymbol\Omega$ so that we can assess the relationship of all taxa within and between the communities. The second is to identify the community labels of all the taxa given by the parameter $\boldsymbol{z}$. Thirdly, we would like to infer the optimal value of $K$ to estimate the appropriate number of communities. 

Bayesian inference commonly uses simple numerical summaries such as the posterior mean to obtain a point estimate of a model parameter \citep{hoff2009first}. The point estimate for each $\omega_{kk'}\in\boldsymbol\Omega$ is computed as the posterior mean of the after burn-in posterior samples, which is given by
\begin{equation}
    \hat{\omega}_{kk'}=\frac{1}{T-B}\sum_{t=B+1}^T\omega_{kk'}^{(t)}
\end{equation}
where $t$ is the current iteration of the MCMC algorithm after a burn-in period.

Next, we infer the community labels of all $p$ taxa \textit{via} the parameter $\boldsymbol{z}$.  We could identify the $\boldsymbol{z}^{(t)}$ at a particular iteration $t$ (after burn-in) that maximizes the posterior distribution. This is known as the \textit{maximum a posteriori} (MAP) estimate and is denoted as $ \hat{z}^{\text{MAP}}$.  Specifically,
\begin{equation}
    \hat{z}^{\text{MAP}} = \argmax_{\boldsymbol{z}\in\{\boldsymbol{z}^{(B+1)},\ldots,\boldsymbol{z}^{(T)}\}}\pi(\boldsymbol{z},\boldsymbol\Omega|\boldsymbol{G},\boldsymbol Q).
\end{equation}
%where the superscripts $(B+1),\ldots,(T)$ identify the posterior samples after a burn-in period and $\pi(\boldsymbol\Omega,\boldsymbol{z}|\boldsymbol{G})$ denotes the joint posterior distribution. 

Bayesian information criterion (BIC) \citep{schwarz1978estimating} is one popular metric for determining the optimal number of communities for model-based clustering algorithms \citep{silva2019multivariate, mcnicholas2016mixture}. BIC is defined as 
\begin{equation}
\text{BIC}=\nu\log{p}-2\left(\argmax_{\pi(\boldsymbol{z},\boldsymbol\Omega|\boldsymbol{G},\boldsymbol Q)}\log\pi(\boldsymbol{z},\boldsymbol\Omega|\boldsymbol{G},\boldsymbol Q)\right)
\end{equation}
where $\nu$ is the number of model parameters and $p$ is the number of taxa. The number of parameters is $\nu=1+K(K+1)/2$ since we have to estimate  $\boldsymbol{z}$ and $K(K+1)/2$ edge probabilities given by each $\omega_{kk'}\in\boldsymbol{\Omega}$. BIC is an appropriate metric when the sample size $n$ is greater than the number of model parameters \citep{giraud2021introduction}, which was not an issue for all analyses in Section \ref{results}.

\section{Results}\label{results}
In this section, we describe the generative model in our simulation study and demonstrate the superior performance of {Bayesian-SBM-MRF} through both simulation and two case studies of very different kinds. 

\subsection{Simulation}\label{sbm_simulation2}
We constructed co-occurrence networks with species-level taxa as nodes that are grouped into $K$ communities. %Two taxa $j\in k$ and $j'\in k'$ may belong the same community ($k=k')$ or two different communities ($k\neq k')$. The existence of an edge between two taxa has probability $\omega_{kk'}.$ 
In particular, we sampled each edge of the network as $g_{jj'}|\omega_{kk'}\sim\text{Bernoulli}(\omega_{kk'})$. To incorporate the taxonomic tree information, we randomly assigned each of the $p$ species to a community and genus using three strength settings (weak, moderate, and strong). Strength here describes how informative the taxonomic tree information can be for community detection. Let $\boldsymbol{\tau}=(\tau_1,\ldots,\tau_p)^\top$ specify the genus labels for species $1,\ldots,p$. We used the adjusted Rand index (ARI) \citep{rand1971objective}, which is a similarity metric between two sets of discrete labels (e.g., community and genus), to classify strength. Equation \eqref{ariequation} below shows how to compute ARI, which usually takes on values between $0$ and $1$. An ARI close to zero indicates little to no similarity between community and genus, and closer to one indicates strong similarity. We considered the tree information to be weakly informative if the ARI between the genus and community labels was low (e.g., $\text{ARI}(\boldsymbol{z},\boldsymbol\tau) \leq 0.3)$. Next, the tree information was moderately informative if $0.3< \text{ARI}(\boldsymbol{z},\boldsymbol\tau) \leq 0.7$ and strongly informative if $0.7<\text{ARI}(\boldsymbol{z},\boldsymbol\tau) \leq 1$. Table \ref{tab:tabari} gives examples of the weak, moderate, and strong settings for ten taxa ($j=1,\ldots,10$) and two communities ($z_j\in\{1,2\}$). Each genus $\tau_j$ is a natural number. Under the weak setting where $\text{ARI}(\boldsymbol{z},\boldsymbol\tau)=0$, none of the taxa belong to the same genus, and so there is no similarity between community and genus. The moderate setting where $\text{ARI}(\boldsymbol{z},\boldsymbol\tau)=0.5$ illustrates a fair level of similarity between the genus and community where the taxa from the first community all belong to the same genus; however, the taxa in the second community all belong to distinct genera. The strong setting where $\text{ARI}(\boldsymbol{z},\boldsymbol\tau)=1$ illustrates the perfect scenario where taxa belonging to the first community all belong to the same genus and taxa in the second community all belong to another genus. Additionally, diversity can be inferred from the strength of the taxonomic tree information. As the level of strength weakens, the diversity in genus increases. Similarly, diversity decreases and strength increases.
\begin{table}[h!]
    \footnotesize
        \caption{Examples of weak, moderate, and strong simulation settings with $p=10$ taxa and $K=2$ communities. $\tau_j$ and $z_j$ give the genus and community of taxa $j$, respectively, for $j=1,\ldots,10$. The quantity $\text{ARI}(\boldsymbol\tau,\boldsymbol{z})$ gives the ARI between genus and community.}
    \label{tab:tabari}
    \centering
    \begin{tabular}{lcccccccccc}
    \toprule
    \multicolumn{1}{c}{}&\multicolumn{10}{c}{Taxon $j$}  \\
    \cmidrule{2-11}
    \multicolumn{1}{c}{}&\multicolumn{1}{c}{1}&\multicolumn{1}{c}{2}&\multicolumn{1}{c}{3}&\multicolumn{1}{c}{4}&\multicolumn{1}{c}{5}&\multicolumn{1}{c}{6}&\multicolumn{1}{c}{7}&\multicolumn{1}{c}{8}&\multicolumn{1}{c}{9}&\multicolumn{1}{c}{10}\\\midrule
   \multicolumn{1}{c}{}& \multicolumn{10}{c}{\underline{Weak Setting}} \\
    $\tau_j$ (Genus) & 15 & 10 & 17 & 6 & 11 & 9 & 25 & 29 & 3 & 1\\
    $z_j$ (Community) & 1 & 1 & 1 & 1 & 1 & 2 & 2 & 2 & 2 & 2 \\
    \multicolumn{1}{c}{}& \multicolumn{10}{c}{$\text{ARI}(\boldsymbol\tau,\boldsymbol{z})=0$} \\
      \multicolumn{11}{c}{} \\
      \multicolumn{1}{c}{}&\multicolumn{10}{c}{\underline{Moderate Setting}} \\
       $\tau_j$ (Genus) & 15 & 15 & 15 & 15 & 15 & 3 & 4 & 26 & 7 & 8\\
    $z_j$ (Community) & 1 & 1 & 1 & 1 & 1 & 2 & 2 & 2 & 2 & 2 \\
    \multicolumn{1}{c}{}& \multicolumn{10}{c}{$\text{ARI}(\boldsymbol\tau,\boldsymbol{z})\approx 0.5$} \\
       \multicolumn{11}{c}{} \\
      \multicolumn{1}{c}{}&\multicolumn{10}{c}{\underline{Strong Setting}} \\
       $\tau_j$ (Genus) & 15 & 15 & 15 & 15 & 15 & 8 & 8 & 8 & 8 & 8\\
    $z_j$ (Community) & 1 & 1 & 1 & 1 & 1 & 2 & 2 & 2 & 2 & 2 \\
    \multicolumn{1}{c}{}& \multicolumn{10}{c}{$\text{ARI}(\boldsymbol\tau,\boldsymbol{z})=1$} \\
     \bottomrule
    \end{tabular}
\end{table}

Next, we simulated co-occurrence networks that are composed of $K=\{3,6,9\}$ communities and $p=180$ species-level taxa. Edge probabilities between communities are typically low, so we randomly sampled these edges from a uniform distribution such that $\omega_{kk'}\sim\text{Uniform}(0,0.1)$ for $k\neq k'$. The probability of an edge between two taxa within the same community $\omega_{kk'}$ where $k=k'$ took on preset values between $0$ and $1$ to imitate various levels of taxon-taxon interaction in each community. Specifically, $\omega_{kk'}=\{0.3,0.6,0.95\}$ for $k=1,\ldots,3$ (i.e., $K=3$), $\omega_{kk'}=\{0.1,0.3,0.5,0.7,0.9,0.97\}$ for $k=1,\ldots,6$ (i.e., $K=6$), and $\omega_{kk'}=\{0.12,0.2,0.3,0.4,0.5,0.7,0.8,0.9,0.99\}$ for $k=1,\ldots,9$ (i.e., $K=9$).  We included $30$ genera in total where $\tau_j\in\{1,\ldots,30\}$ for each taxon $j=1,\ldots,180$. Altogether, there were $3\times 3=9$ settings in this simulation. For each scenario, we repeated the above steps to generate $50$ replicates. 

First, we wanted to determine if \text{Bayes-SBM-MRF} improves the performance of the standard Bayesian SBM by incorporating taxonomic tree information. We compared the results of {Bayesian-SBM-MRF} under four settings of the MRF prior when $f=\{0,0.5,1,2\}$. When $f=0,$ the standard Bayesian SBM is employed since no taxonomic tree information is incorporated. This setting was then compared to the other three when $f>0$ where \text{Bayes-SBM-MRF} incorporates taxonomic tree information. Second, we compared these four settings to two commonly used competing clustering models selected from Table \ref{tab:sbm_software}: the \texttt{cluster\_fast\_greedy} function from the very popular \texttt{igraph} package in \texttt{R}, and spectral clustering from the \texttt{anocva} package in \texttt{R}. 

We assessed the performance of all models on the $50$ replicated data sets under each setting to determine how well they can recover the true community labels $\boldsymbol{z}$. One popular similarity metric for comparing the predicted and true community labels to assess model quality is ARI, which is useful for dealing with the issue of community label switching that is common to clustering algorithms \citep{jiang2020bayessmiles}. ARI is a corrected version of the Rand index so that label switching is not an issue when comparing predicted and true community labels. For example, an algorithm may estimate the community labels of four taxa to be $\boldsymbol{\hat{z}}=\{1,1,2,2\}$ while the underlying truth may be $\boldsymbol{z}=\{2,2,1,1\}$. Both sets of labels are equivalent because the first two taxa are assigned to the same community and the other two taxa are assigned to the other community. Thus, the community label does not matter since it is a nominal-level variable. ARI is calculated as 
\begin{equation}\label{ariequation}
    \text{ARI}(\bm{z}, \hat{\bm{z}}) \quad=\quad \frac{{p \choose 2}(a + d) - [(a + b)(a + c) + (c + d)(b + d)]}{{p \choose 2}^2 - [(a + b)(a + c) + (c + d)(b + d)]}
\end{equation}
 where $a = \sum_{j > j'}\bm{I}(z_j = z_{j'})\bm{I}(\hat{z}_j = \hat{z}_{j'})$, $b = \sum_{j > j'}\bm{I}(z_j = z_{j'})\bm{I}(\hat{z}_j \neq \hat{z}_{j'})$, $c = \sum_{j > j'}\bm{I}(z_j \neq z_{j'})\bm{I}(\hat{z}_j = \hat{z}_{j'})$ and $d = \sum_{j > j'}\bm{I}(z_j \neq z_{j'})\bm{I}(\hat{z}_j \neq \hat{z}_{j'}),$ respectively. ARI is usually bounded between zero and one, but it may take on small negative values. An ARI closer to one indicates greater similarity between the predicted and true community labels, which provides evidence that the model can recover the underlying communities of a microbiome co-occurrence network.

\begin{figure}
    \centering
    \includegraphics[width=1.0\linewidth]{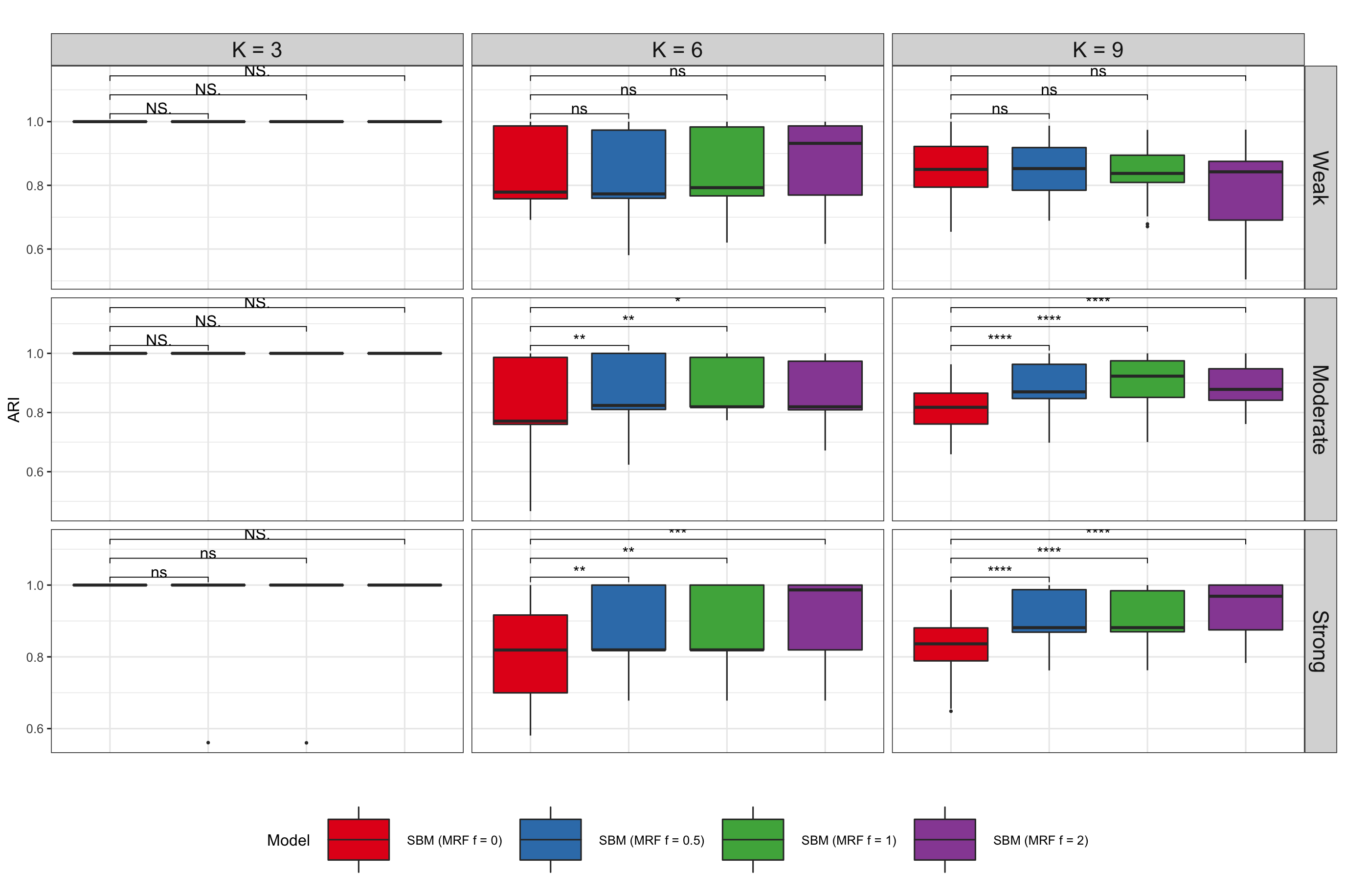}
    \caption{Boxplots of adjusted Rand index (ARI) with four settings of the MRF prior ($f=\{0,0.5,1,2\}$) for the simulated data sets under nine settings of combined number of communities $K=\{3,6,9\}$ and the level of informative strength of the taxonomic tree information \{Weak, Moderate, Strong\}. The paired Mann-Whitney-Wilcoxon test was performed to compare the prior setting where $f=0$ to the other settings where $f>0$ to assess model performance with and without taxonomic tree information. Significance is indicated by * ($p<0.05$), ** ($p<0.01$), *** ($p<0.001$), **** ($p<0.0001$), ns (not significant with $p\in(0.05,1)$), NS (not significant with $p=1$).}
    \label{fig:ari_mrf_only}
\end{figure}

\begin{figure}
    \centering
    \includegraphics[width=1.0\linewidth]{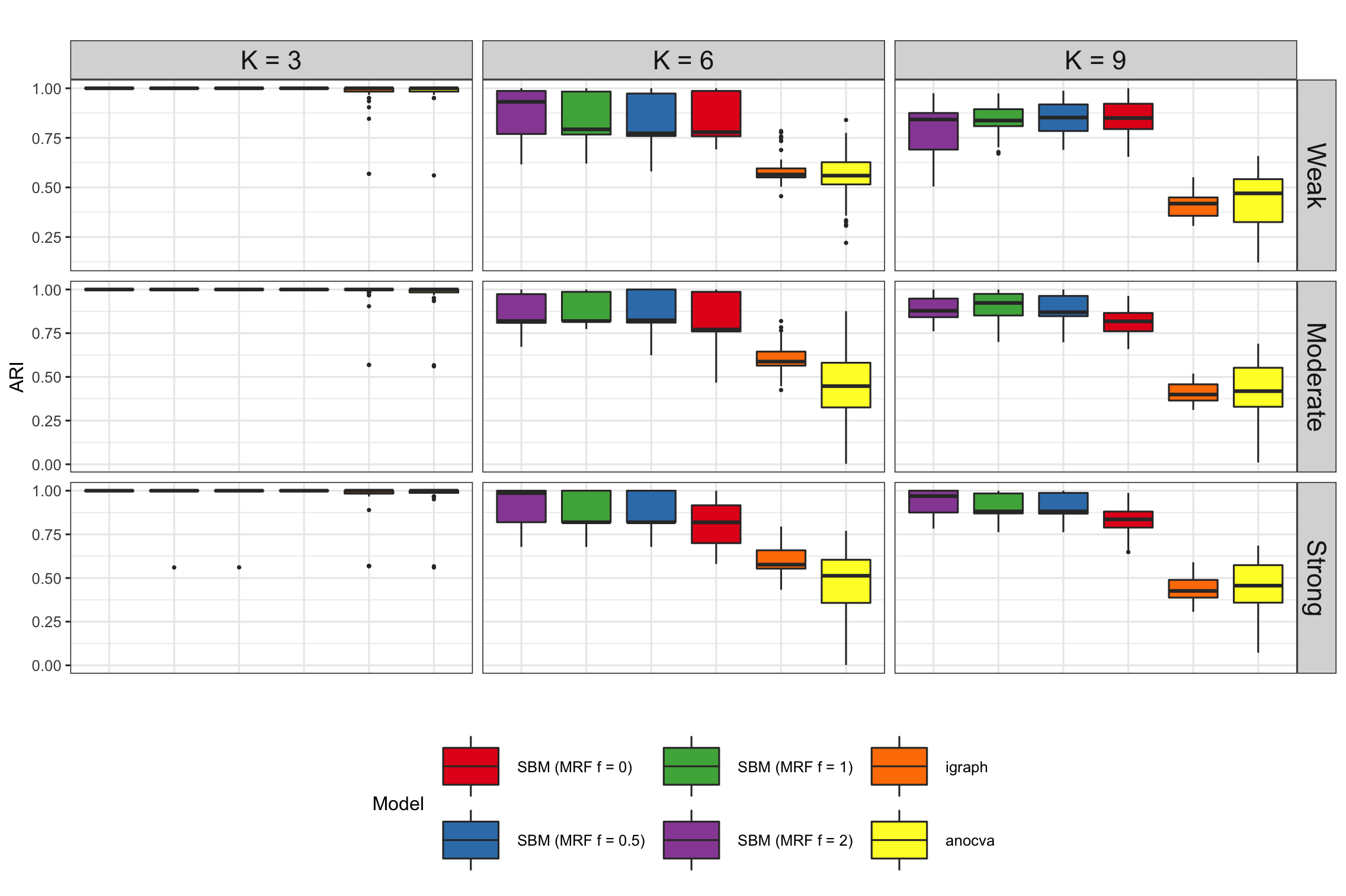}
    \caption{Boxplots of adjusted Rand index (ARI) with four settings of the Markov random field prior ($f=\{0,0.5,1,2\}$) compared to two competing methods: \texttt{cluster\_fast\_greedy} function from the \texttt{igraph} library in \texttt{R}, and spectral clustering from the \texttt{anocva} library in \texttt{R}. The simulated data sets had nine combined settings of the number of communities $K=\{3,6,9\}$ and the level of informative strength of the taxonomic tree information \{Weak, Moderate, Strong\}.}
    \label{fig:ari_competing}
\end{figure}

Results comparing the ARI of the four settings of \text{Bayes-SBM-MRF} are displayed in Figure \ref{fig:ari_mrf_only}. ARI was computed using the \texttt{ARI} function from the \texttt{aricode} library in \texttt{R}.
We performed the paired Mann-Whitney-Wilcoxon test to compare the standard Bayesian SBM when $f=0$ to the other three settings of our generalized model. Several important observations stand out about our generalized SBM model. The first is that when the level of strength of the taxonomic tree information is moderate or high, \text{Bayes-SBM-MRF} where $f>0$ has significantly higher ARI than the standard Bayesian SBM where $f=0$. This indicates that the taxonomic tree information can be useful and more powerful for recovering the true community labels, especially when it is informative. Second, there are no significant differences between all four model settings when strength is weak, which indicates that the inclusion of taxonomic tree information does not hurt the performance of our generalized model even when it is not informative. Third, we recommend using $f=1$ for \text{Bayes-SBM-MRF} because it performs equally or better than the other settings. When $K=9$ and strength is weak, the boxplot corresponding to $f=2$ is very negatively skewed, indicating the start of a possible phase transition.  
Next, the boxplots of the ARI for all competing models in Figure \ref{fig:ari_competing} demonstrate that {Bayesian-SBM-MRF} has superior performance under all nine settings. Also, {Bayesian-SBM-MRF} has consistent ARI as $K$ increases; whereas, the ARI of the competitors decreases as $K$ increases. This implies that the competing methods are unreliable when there may be many underlying communities. Our method is consistent for both small and large values of $K$.

\subsection{Real Data Analysis 1: Urinary Microbiome Data}\label{joint_data_analysis}

We applied {Bayesian-SBM-MRF} to the urinary microbiome data from our study on rUTI with $n=75$ postmenopausal female patients. This is the first time anyone has studied the network and community structure of the urinary microbiome with respect to rUTI in postmenopausal women. The urinary microbiome data originally had $180$ bacterial species along with their known genera. We filtered out all species that had fewer than seven non-zero counts, which resulted in a total of $p=99$ species from $41$ genera. Of the $41$ genera, $18$ of these had at least two species belonging to the same genus. About $77\%$ of the bacterial species belong to these $18$ genera. Each of the remaining $23\%$ of species belongs to unique genera. The full details regarding the study, data, and metagenomic sequencing can be found in \citet{neugent2022recurrent}.

The microbiome co-occurrence network $\boldsymbol{G}$ was estimated from the species-level abundance data and the available taxonomic tree information was incorporated into $\boldsymbol{Q}$ as described in Section \ref{sbm_data}. The MCLR transformation was done using the \texttt{MCLR} function from the \texttt{SPRING} library in \texttt{R}. Then, we performed community detection on $\boldsymbol{G}$ using {Bayesian-SBM-MRF} with $f=1$ for the MRF prior setting. We set $\eta_k=\log(1/K)$ to impose a non-informative discrete uniform prior on each community label $z_j\in\boldsymbol{z}$. We set $a_\omega=b_\omega=1$ to impose a non-informative uniform prior on each edge probabilities $\omega_{kk'}\in\boldsymbol{\Omega}$. We ran $T=1000$ iterations, with the first half of the iterations discarded as burn-in samples.

Our {Bayesian-SBM-MRF} selected $K=7$ communities \textit{via} BIC as the optimal number of communities (see the elbow plot in Figure \ref{fig:bic_urinary} in the appendix). Figure \ref{fig:data_heatmap} is the heatmap of the estimated microbiome co-occurrence network organized into the seven communities determined by {Bayesian-SBM-MRF} with $K=7$. Names of genus and species are in the margins along with their community label number. The main diagonal runs from bottom-left to top-right and represents the seven communities along with the taxa that interact within those communities. These seven communities on the main diagonal are arranged in order from least to greatest within-community edge probabilities (i.e., darker indicates higher edge probability). The off-diagonal blocks represent between-community interactions. As we can see, communities 6 and 7 at the top right of the main diagonal are equally as dark, indicating that they have about the same edge probability. Even though they are very similar in that sense, the model distinguishes them because their interactions with other communities are very different. The $7^{\text{th}}$ community has high interaction with communities 4 and 5; whereas, community 6 barely interacts with any other community. Even though we imposed a discrete uniform prior on the community labels, notice that the community sizes are quite different. This seems to indicate that there is a good balance of information coming from both the microbiome co-occurrence network and the taxonomic tree information to inform community detection. 
\begin{figure}
    \centering
    \includegraphics[width = 1.0\linewidth]{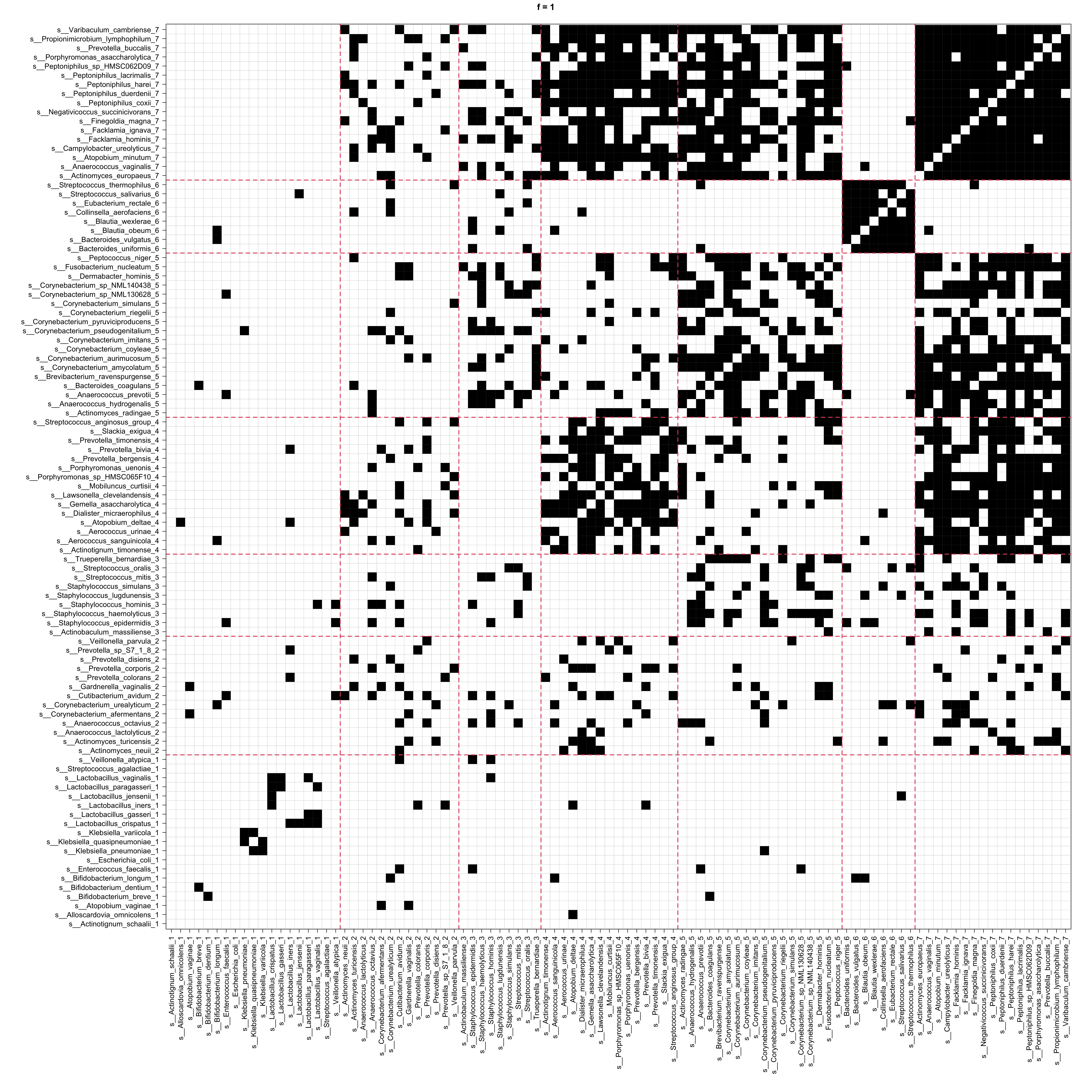}
    \caption{A heatmap of $K=7$ communities detected by {Bayesian-SBM-MRF} ($f=1$). Names of genus and species are in the margins along with their community label number. The main diagonal represents the seven communities determined by the model and their within-community interactions. Off-diagonal blocks represent the between-community interactions. Black indicates a significant association between two taxa and white indicates otherwise.}
    \label{fig:data_heatmap}
\end{figure}

\begin{figure}
    \centering
    \includegraphics[width = 1.0\linewidth]{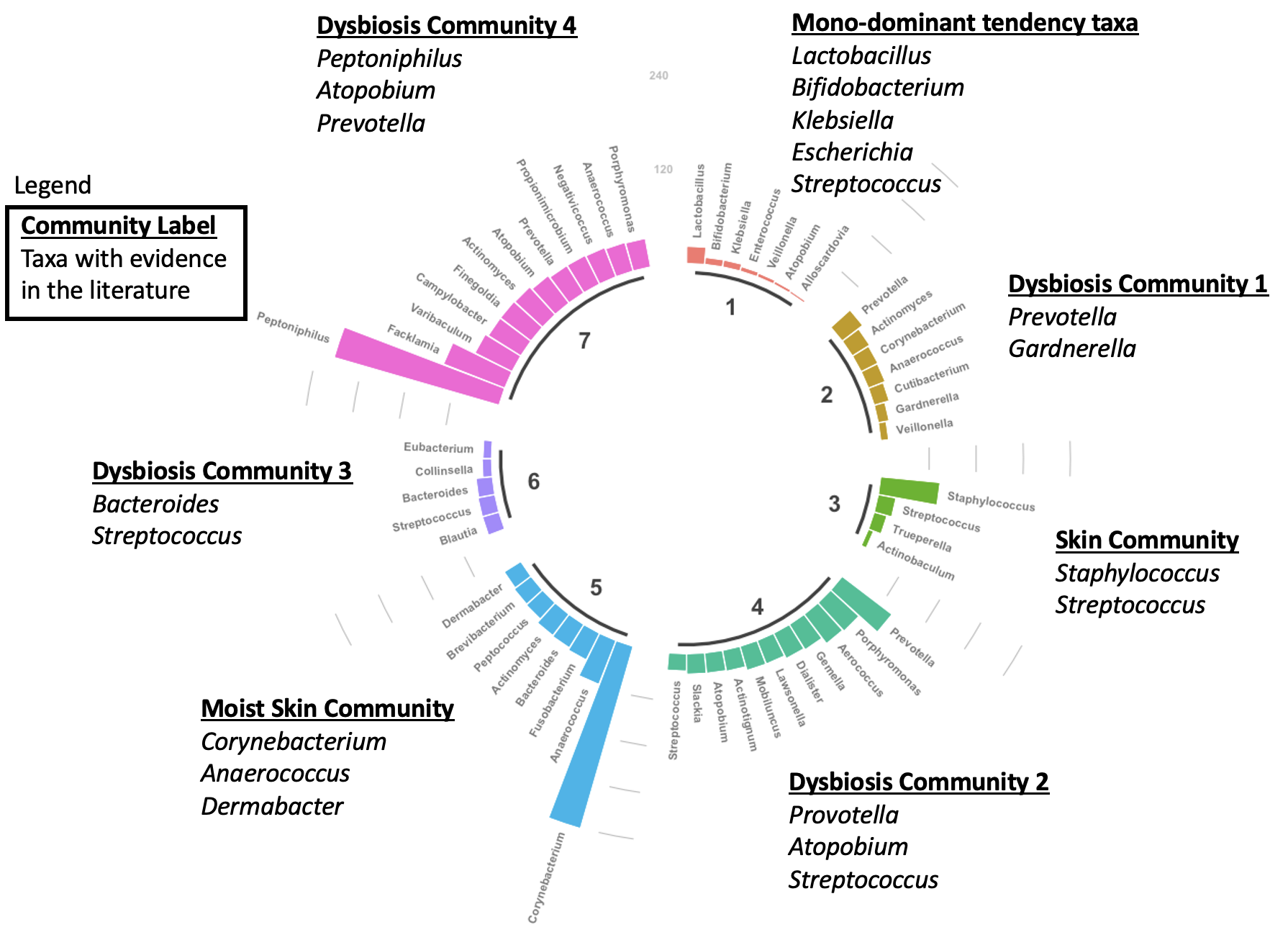}
    \caption{A circular bar plot of $K=7$ communities detected by {Bayesian-SBM-MRF} ($f=1$). Labels on the bars identify the genera found in each community. Bar frequency gives the nodal strength of each genus per community. Annotations identify genera with evidence-based characteristics.}
    \label{fig:annotated_circle_bar plot}
\end{figure}

We calculated the nodal strength with respect to genus in each community as a way to compare and classify the detected communities. Nodal strength is defined as the sum of all edge weights of a single taxon \citep{hall2019co}. Nodal strength for taxon $j$ is denoted as $d_j$ and is computed as  
\begin{equation}
    d_j = \sum_{j'=1}^p I(g_{jj'} = 1).
\end{equation}
Since $\boldsymbol{G}$ is binary, nodal strength is simply nodal degree. Next, we summed each $d_j$ for all taxa in the same community belonging to the same genus to calculate the genus-level nodal strength within each community. This can help to compare both community and network structures. The circular bar plot in Figure \ref{fig:annotated_circle_bar plot} displays the genus-level nodal strength (as frequency) in each of the seven communities. The communities and the genera within those communities have evidence-based characteristics that are of biological interest. These characteristics are noted in the plot annotations next to each respective community. For example, communities 3 and 5 are associated with the human skin microbiome \citep{chen2018skin}. In communities 2, 4, 6, and 7, we also found four distinct microbiome communities harboring taxa known to be markers of dysbiosis in the female urinary microbiome from genera such as \textit{Gardnerella, Prevotella, Streptococcus}, and \textit{Peptoniphilus} \citep{neugent2022recurrent,ceccarani2019diversity, shipitsyna2013composition}.

%For example, four of the groups contain genera with evidence in the literature that certain taxa within those genera are associated with dysbiosis in other microbiomes.  

We also compared the results of standard SBM (i.e., when $f=0$; genus information is not used) to the results above. Table \ref{tab:genus_table} gives some promising insight into our model's behavior. When $f=1,$ we found at least six genera that appear in fewer communities than when $f=0.$  While the genera \textit{Peptoniphilus}, \textit{Staphylococcus}, and \textit{Facklamia} collapsed into one community, other genera did not follow the same pattern. \textit{Atopobium} collapsed from three to two communities, \textit{Corynebacterium} collapsed from four to two communities, and \textit{Anaerococcus} collapsed from four to three communities. All other genera showed no changes in the number of assigned communities. Thus, {Bayesian-SBM-MRF} does not simply force species from the same genus into one community. These observations indicate that {Bayesian-SBM-MRF} is well-behaved because it is likely making a balanced use of both the microbiome co-occurrence network $\boldsymbol{G}$ and the taxonomic tree information $\boldsymbol{Q}$. The remaining analysis focuses on the difference in diversity of genera per community for $f=0$ versus $f=1$.

\begin{table}[h!]
    \centering
     \caption{The number of communities that select genus have been assigned to under two different settings ($f=0$ and $f=1$) of the generalized model.}
    \label{tab:genus_table}
    \begin{tabular}{lcc}
    \hline
      \textbf{Genus}   & $f=0$ & $f=1$ \\\hline
       \textit{Peptoniphilus}   & 2 & 1 \\
        \textit{Facklamia} & 2 & 1 \\
        \textit{Bifidobacterium} & 2 & 1 \\      
       \textit{Staphylococcus} & 2 & 1 \\
       \textit{Atopobium} & 3 & 2 \\
       \textit{Corynebacterium} & 4 & 2 \\
       \textit{Anaerococcus} & 4 & 3 \\\hline       
    \end{tabular}
\end{table}

\begin{figure}
    \centering
    \includegraphics[width=1.0\linewidth]{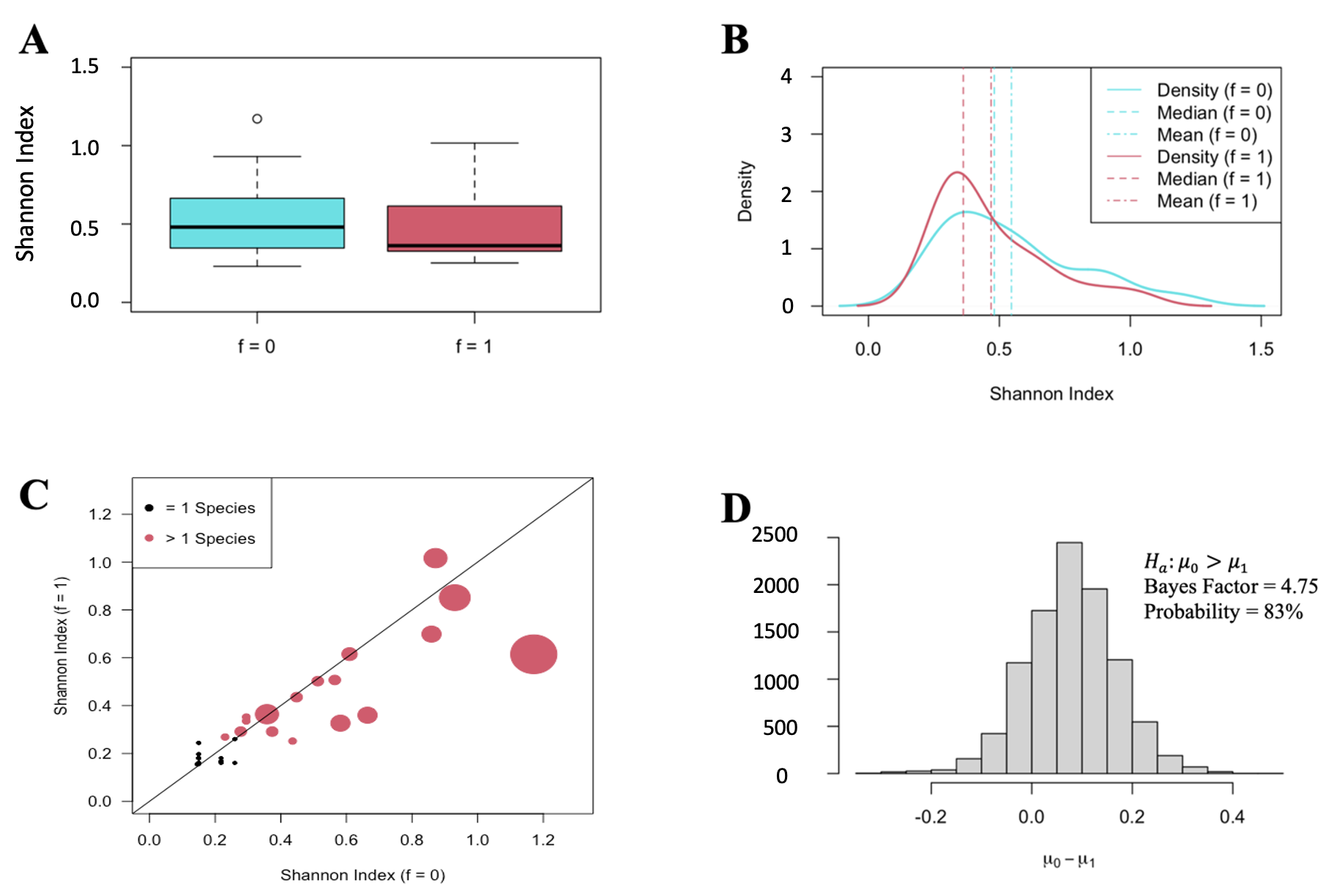}
    \caption{Plots of the Shannon index of the 41 genera pertaining to the standard Bayesian SBM ($f=0$) and {Bayesian-SBM-MRF} ($f=1$). \textbf{A}: Boxplots of the Shannon index comparing the diversity of the two models; \textbf{B}: Density plots; \textbf{C}: Scatter plot of Shannon index of $f=0$ versus $f=1$; the size of points varies with respect to the number of species \textbf{D}: Histogram of the estimated posterior distribution of the difference in Shannon index means of the two models, Bayes factor and posterior probability in favor of the alternative hypothesis.}
    \label{fig:shannon_summary}
\end{figure}

Figure \ref{fig:both_sbm_f01} in the appendix illustrates the difference in community structure using both within-community and between-community nodal strength with respect to genus for both the standard Bayesian SBM and {Bayesian-SBM-MRF} ($f=0$ and $f=1$, respectively). The figure suggests that there is some difference in diversity within several communities while a few other communities display little to no difference with respect to genus. To quantify the diversity of genera in each community, we further calculated the Shannon index for both models. The Shannon index is the log transformation of the weighted geometric mean of proportional species per genus in a given community \citep{tucker2017guide}. In general, the Shannon index $H$ is calculated as
\begin{equation}
    H=-\sum_{r=1}^R w_r\log(w_r)
\end{equation}
where $R$ is the number of genera and $w_r$ is the proportion of species belonging to the $r^{th}$ genus in a given community. Figure \ref{fig:shannon_summary} compares the Shannon indices of the two model settings when $f=0$ and $f=1$. The boxplots (Figure \ref{fig:shannon_summary}A) compare the median diversity. The density plots (Figure \ref{fig:shannon_summary}B) illustrate the means and medians of the Shannon index distributions for each setting. The scatterplot (Figure \ref{fig:shannon_summary}C) illustrates the diversity of the $41$ genera for $f=1$ versus $f=0$ where the size and color of each point is based on the number of species per genus (i.e., nodal strength). The points tend to fall below the $45^\circ$ line indicating lower diversity in the communities when $f=1.$ We used Bayesian hypothesis testing to determine if there is any significant difference between the Shannon index means of the two models. We assumed a normal model with unknown mean and variance. The histogram (Figure \ref{fig:shannon_summary}D) of the difference in the Shannon index means $\mu_0-\mu_1$ for $f=0$ versus $f=1$ is given. The hypotheses of interest are $H_0:\mu_0=\mu_1$ versus $H_a:\mu_0>\mu_1$ since we wish to determine if the mean diversity is significantly lower when $f=1$. We made inferential decisions based on Bayes Factor and the posterior probability in favor of the alternative hypothesis. According to Jeffreys' rule \citep{kass1995bayes}, a Bayes factor between $3.2$ and $10$ indicates substantial evidence to reject the null hypothesis. Here, the Bayes factor is $4.75$ with $83\%$ posterior probability in favor of the alternative hypothesis over the null hypothesis. The use of taxonomic tree information reduces the diversity in the communities, indicating that the results of the standard Bayesian SBM and our {Bayesian-SBM-MRF} are substantially different. %But, the Bayes factor and posterior probability are not impressively significant indicating that the difference between results is not drastic, but still different. 
%These results indicate that {Bayesian-SBM-MRF} may make fair and balanced use of both the taxonomic tree information and the microbiome co-occurrence network. 

\subsection{Real Data Analysis 2: \textit{Les Mis\'erables} Data}\label{miserables}
Co-occurrence networks have been found to be particularly useful in analyzing large text. Here, we demonstrate that our method can be applied to a co-occurrence network derived from a large and structured collection of text.

\textit{Les Mis\'erables} is a French historical fiction novel written by Victor Hugo in the nineteenth century. The novel is comprised of $1462$ pages with $77$ characters. The main character, Jean Valjean, is a loner who is on the run most of his life because he broke his parole after being released from a $19$-year prison sentence for stealing bread to feed his starving family. The story focuses on his interactions with characters throughout his life, such as Javert (the policeman who spends his life hunting down Valjean for breaking his parole), Myriel (the bishop who gives Valjean shelter for the night; gives Valjean a second chance at freedom after he is caught stealing from the bishop's home), Fantine (the woman who on her death bed entrusts her young daughter Cosette into Valjean's care), Monsieur and Madame Th\'enardier (the thieving innkeepers who were the caretakers of Cosette until Valjean comes along), Marius (a student of the French Revolution who falls in love with Cosette; an older Valjean saves this young man's life during the revolution so that he can marry Cosette), and Eponine (who is secretly in love with Marius and is the daughter of the Th\'enardier's). The story also includes tertiary characters such as prisoners, prostitutes, students, servants, etc.

The \textit{Les Mis\'erables} data \citep{knuth1993stanford} are an undirected graph $\boldsymbol{G}$ of $p=77$ nodes. Each node represents a character. An edge is assigned between any two characters if they appear in the same chapter. Edges were also weighted based on the number of co-appearances of any two characters in the same chapter. We utilized the edge weights to construct matrix $\boldsymbol{Q}$ for the MRF prior. Major characters should have a higher frequency of interaction in any given chapter. Tertiary characters, or minor characters, will have lower frequency. So, we decided that if two characters' edge weight was greater than two, then $q_{jj'}=1$ and zero otherwise.  Thus, the matrix $\boldsymbol{Q}$ helps to distinguish tertiary and non-tertiary characters. The network $\boldsymbol{G}$ cannot distinguish character types because all existing edges are equally weighted since they are binary. We ran {Bayesian-SBM-MRF} with the same settings as the urinary microbiome data analysis.

The model selected $K=6$ communities \textit{via} BIC when $f=1$ (see the elbow plot in Figure \ref{fig:bic_lesmis} in the appendix). Figure \ref{fig:lesmis_network} illustrates the six communities. This plot was generated in Gephi version 0.10.1 \citep{bastian2009gephi}. The size of a node was determined by nodal strength, just as in the previous analysis. The most interesting result is that Valjean is in a community by himself, which makes sense in the context of the story since Valjean is the primary character portrayed as a loner and having brief interactions with most characters. The next community of interest is the blue dots, which includes secondary characters with large roles such as the Thenardier's, Javert, Eponine, and Cosette. The green community includes the young students and Gavroche who almost all died in the French Revolution. The orange community includes mostly prostitutes and Fantine, the mother of Cosette, who dies early in the story from tuberculosis shortly after having no choice but to become a prostitute herself to support her child. The purple community contains the judge and criminals. Valjean is in prison with these criminals at the beginning of the novel. One of the prisoners, Bamatabois, also interacts with Fantine by viciously demanding her services. Lastly, the yellow dots are tertiary characters (e.g., Child 1, Child 2, Woman 1, Woman 2), who appear very briefly throughout the novel. Lastly, we analyzed these data using the standard Bayesian SBM ($f=0$) for the sake of comparison. All characters remained in the same communities except for Cosette. Cosette was placed into the community of tertiary characters, which makes no sense since she is a main character. By including additional information \textit{via} the MRF prior, \text{Bayes-SBM-MRF} returned a slightly different but more sensible result than the standard Bayesian SBM. %In summary, {Bayesian-SBM-MRF} grouped the characters into communities that makes sense with respect to the timeline and character relationships when $f=1$ indicating that our model does a very good job at performing community detection. 
This demonstrates that {Bayesian-SBM-MRF} does a better job at performing community detection than the standard Bayesian SBM and can be broadly applied to other types of network data. 

\begin{figure}
    \centering
    \includegraphics[width=1.0\linewidth]{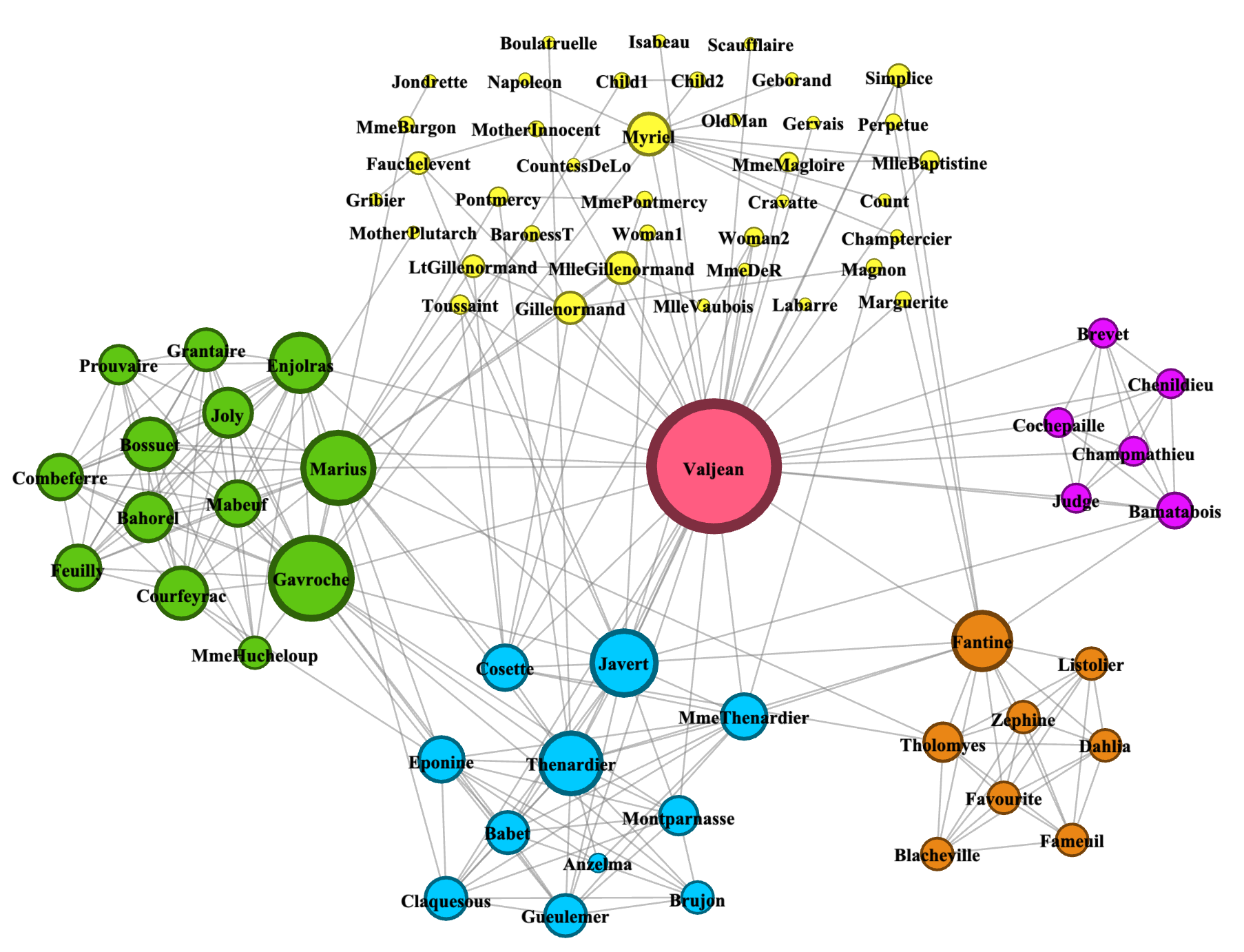}
    \caption{The network of characters from \textit{Les Mis\'erables} and the six communities determined by {Bayesian-SBM-MRF}. Each color denotes a distinct community. Each edge indicates that two characters appear in the same chapter at least once. Nodal size was determined by nodal strength.}
    \label{fig:lesmis_network}
\end{figure}

\section{Discussion}\label{conclusion}

In this article, we develop a two-stage method for community detection on a microbiome co-occurrence network. The first stage estimates the microbiome co-occurrence network from the MCLR-transformed relative abundances to account for their compositionality, non-linearity, and zero-inflation. We believe this is the first time the MCLR transformation has been used for community detection. The second stage takes the estimated microbiome co-occurrence network and available taxonomic tree information into account to perform community detection using a generalized SBM model, {Bayesian-SBM-MRF}. We believe that {Bayesian-SBM-MRF} is the first of its kind to include two levels of the taxonomic tree to perform community detection. 
The simulation study revealed several advantages of {Bayesian-SBM-MRF}. First, the inclusion of taxonomic tree information improves model performance when it is informative. Second, the inclusion of taxonomic tree information does not hurt model performance even when it is non-informative. Third, {Bayesian-SBM-MRF} demonstrated superior performance over commonly used competing methods in all simulation settings. When applied to a real urinary microbiome data set from a study on rUTI from postmenopausal women, {Bayesian-SBM-MRF} uncovered several communities characterized by evidence-based dysbiosis.  Further, this was the first time community detection has been done to study the urinary microbiome co-occurrence network and community structure with respect to rUTI from postmenopausal women. Our findings provide the foundation for future studies in this particular area of research. Additional studies will be required to validate the biological relevance of these communities and their association with rUTI. The application of {Bayesian-SBM-MRF} to the \textit{Les Mis\'erables} data demonstrates that our model can also be broadly applied to other types of undirected networks. {Bayesian-SBM-MRF} is more appropriate for microbiome co-occurrence networks because a stochastic model can handle a high-dimensional network; however, it provides reasonable results for other applications and smaller networks. 

There are several limitations of {Bayesian-SBM-MRF}. As mentioned above, our method accounts for zero-inflation in the abundance data using an external normalization step (i.e., the MCLR transformation). Other methods, such as HARMONIES \citep{jiang2020harmonies} and SPRING \citep{yoon2019microbial}, proposed zero-inflated model-based methods for estimating a network using probability distributions such as the zero-inflated negative binomial distribution. Both methods demonstrated superior performance over well-known network analysis methods using model-based or internal normalization. {Bayesian-SBM-MRF} is a multi-stage model that first estimates the microbiome co-occurrence network and then performs community detection. Previous studies have demonstrated that a joint model is more efficient and may have superior performance when compared to multi-stage models \citep{leaman2016taggerone,lou2017transition,zhao2019neural,koslovsky2020bayesian,ji2021neural}. There are alternative methods in the literature for determining the optimal number of communities. While BIC is standard and widely used, some issues exist. BIC may be unreliable when the sample size is less than the number of model parameters \citep{giraud2021introduction}.  \citet{hu2020corrected} recently proposed a corrected BIC (CBIC) designed specifically for selecting the optimal value of $K$ in a stochastic block model. They found that standard BIC may overestimate the true number of communities. CBIC adds an additional penalty to the log-likelihood, which may better estimate the optimal value of $K.$ \citet{biernacki2000assessing} proposed the integrated complete likelihood (ICL) to determine the optimal value of $K$ and also found that standard BIC may overestimate $K$. Each of these limitations can be explored in future work to help possibly improve model performance.  

%\section*{Acknowledgements}

\section*{Funding Information}
This work was supported by grants 1R01DK131267-01 and 1R01GM141519 from the National Institutes of Health, 2210912 and 2113674 from the National Science Foundation, and AT-2030-20200401 from the Welch Foundation. NJD was supported by a research grant from the Foundation for Women's Wellness. MLN was supported by National Institutes of Health Fellowship 1F32DK128975-01A1. 

\section*{Conflict of Interest}
The authors declare no potential conflict of interest.

\section*{Data Availability Statement}
Whole genome metagenomic sequencing read data (FASTQ files) have been deposited onto the NIH Sequence Read Archive (SRA) under the BioProject number [NCBI]: PRJNA801448. Prior to depositing, all human-mapping reads were removed from the data. Simulated data, filtered real urinary microbiome abundance data used for analysis, and the related source code in \texttt{R} are available at \url{https://github.com/klutz920/Bayesian-SBM-MRF}. The \textit{Les Mis\'erables} data are publicly available in the \texttt{gsbm} library in \texttt{R}.

\section*{Orcid}

    \begin{tabular}{ll}
     \textit{Kevin C. Lutz}    &  \href{https://orcid.org/0000-0002-6687-0637}{https://orcid.org/0000-0002-6687-0637}  \\
     \textit{Michael L. Neugent} & \href{https://orcid.org/0000-0002-9863-9595}{https://orcid.org/0000-0002-9863-9595} \\
      \textit{Tejasv Bedi }   &  \href{https://orcid.org/0000-0001-7532-4075}{https://orcid.org/0000-0001-7532-4075} \\
      \textit{Nicole J. De Nisco}   &  \href{https://orcid.org/0000-0002-7670-5301}{https://orcid.org/0000-0002-7670-5301}   \\
      \textit{Qiwei Li}   &   \href{https://orcid.org/0000-0002-1020-3050}{https://orcid.org/0000-0002-1020-3050} \\
    \end{tabular}

\newpage
\appendix
\section{Appendix of Supplementary Tables and Figures}
\begin{table}[!htbp]
    \centering
        \caption{Available software for stochastic block models in \texttt{R}, \texttt{Python}, and \texttt{C++} languages. Abbreviations: (i) Approach is either Bayesian (Bayes) or frequentist (Freq); (ii) Graph is either directed (D) or undirected (U); (iii) Data can be binary (B), discrete (P), or continuous (C); (iv) Companion publications for some of the packages are currently not available (NA).}
    \label{tab:sbm_software}
    \small
    \begin{tabular}{|l|l|l|l|l|l|l|}
    \hline
         {Package} &  {Approach} &  {Graph} &  {Data} &  {Language} &  {Repository} &  {Publication} \\\hline
         
        \texttt{anocva} & Freq & U & B,P,C & \texttt{R} & CRAN & \citep{vidal2017anocva} \\
        \texttt{BipartiteSBM} & Bayes & U & B & \texttt{Python/C++} & GitHub & NA  \\
        \texttt{blockmodeling} & Freq & U & B,P,C &\texttt{R} & CRAN & \citep{vziberna2007generalized}  \\
        \texttt{blockmodels} & Freq & D,U & B,P,C &\texttt{R}& CRAN & \citep{leger2016blockmodels}  \\
        \texttt{CommunityDetection} & Freq & U & B  & \texttt{Python} & GitHub & NA  \\
        \texttt{dBlockmodeling} & Freq & D,U & B,P,C &\texttt{R} & CRAN & \citep{brusco2021deterministic} \\
        \texttt{dynSBM} & Freq & U & B,P,C &\texttt{R} & CRAN & \citep{matias2017statistical}  \\
        \texttt{expSBM} & Freq & D,U & B,C &\texttt{R} & CRAN & \citep{rastelli2019dynamic}  \\
        \texttt{graphon} & Bayes & U & B &\texttt{R} & CRAN & \citep{orbanz2014bayesian}  \\
        \texttt{graph-tool} & Bayes & D,U & B,P,C & \texttt{Python} & Online & \citep{peixoto2019bayesian}  \\
        \texttt{greed} & Bayes & D,U & B,P,C &\texttt{R} & CRAN & \citep{come2022greed}  \\
        \texttt{GREMLIN} & Freq & D,U & B,P,C & \texttt{R} & CRAN & \citep{bar2020block}  \\
        \texttt{hergm} & Bayes & D,U & B &\texttt{R} & CRAN & \citep{schweinberger2018hergm}  \\
        \texttt{igraph} & Freq & D,U & B,P,C & \texttt{R} & CRAN & \citep{csardi2006igraph} \\
        \texttt{missSBM} & Freq & D,U & B &\texttt{R} & CRAN & \citep{barbillon2019misssbm} \\
        \texttt{MixeR} & Bayes, Freq & U & B,P,C & \texttt{R}& CRAN & NA  \\
        \texttt{MODE-NET} & Freq & U & B & \texttt{C++} & Online & NA  \\
        \texttt{noisySBM} & Freq & D,U & C & \texttt{R} & CRAN & \citep{rebafka2019graph}   \\
        \texttt{pysbm} & Bayes, Freq & D,U & B,P,C & \texttt{Python} & GitHub & \citep{funke2019stochastic}  \\
        \texttt{sbm} & Freq & D,U & B,P,C & \texttt{R} & CRAN & NA  \\
        \texttt{sbm\_canonical\_mcmc} & Bayes & U & B & \texttt{C++} & GitHub & \citep{young2017finite}  \\
        \texttt{sbmr} & Bayes & U & B,P & \texttt{R}& GitHub & NA  \\
        \texttt{sbmSDP} & Freq & U & B &\texttt{R} & CRAN & NA \\
        \texttt{SBMSplitMerge} & Bayes & D,U & B,P,C  &\texttt{R} & CRAN & \citep{ludkin2020inference}  \\
        \texttt{SparseBM} & Freq & D,U & B & \texttt{Python} & GitHub & NA \\
        \hline
    \end{tabular}
\end{table}
\begin{table}[!htbp]
    \centering
        \caption{Online documentation for implementing available stochastic block model software packages. The year of the most recent software update is provided in the last column.}
    \label{tab:sbm_url}
    \begin{tabular}{lll}
    \hline
        Package & URL & Updated\\\hline
        \texttt{anocva} & \url{https://cran.r-project.org/web/packages/anocva/index.html} & 2023 \\
       \texttt{BipartiteSBM}  & \url{https://github.com/junipertcy} & 2020 \\
       \texttt{blockmodeling} & \url{https://cran.r-project.org/web/packages/blockmodeling/index.html} & 2022 \\
       \texttt{blockmodels} & \url{https://cran.r-project.org/web/packages/blockmodels/index.html} & 2021 \\
       \texttt{CommunityDetection} & \url{https://github.com/Jonas1312/community-detection-in-graphs} & 2017 \\
       \texttt{dBlockmodeling} & \url{https://cran.r-project.org/web/packages/dBlockmodeling/index.html} & 2020 \\
       \texttt{dynSBM} & \url{https://cran.r-project.org/web/packages/dynsbm/index.html} & 2020 \\
       \texttt{expSBM} & \url{https://cran.r-project.org/web/packages/expSBM/index.html} & 2019 \\
       \texttt{graphon} & \url{https://cran.r-project.org/web/packages/graphon/index.html} & 2021 \\
       \texttt{graph-tool} & \url{https://graph-tool.skewed.de} & 2017 \\
       \texttt{greed} & \url{https://cran.r-project.org/web/packages/greed/index.html} & 2022 \\
       \texttt{GREMLIN} & \url{https://cran.r-project.org/web/packages/gremlin/index.html} & 2021 \\
       \texttt{hergm} & \url{https://cran.r-project.org/web/packages/hergm/index.html}  & 2021\\
       \texttt{igraph} & \url{https://cran.r-project.org/web/packages/igraph/index.html} & 2023  \\
       \texttt{missSBM} & \url{https://cran.r-project.org/web/packages/missSBM/index.html} & 2022 \\
       \texttt{MixeR} & \url{https://rdrr.io/cran/mixer/} & 2018 \\
       \texttt{MODE-NET} & \url{http://www.lps.ens.fr/~krzakala/MODE_NET/} & 2012 \\
       \texttt{noisySBM} & \url{https://cran.r-project.org/web/packages/noisySBM/index.html} & 2020 \\
       \texttt{pysbm} & \url{https://github.com/funket/pysbm/tree/master/pysbm} & 2019\\
       \texttt{sbm} &\url{https://cran.r-project.org/web/packages/sbm/index.html} & 2021 \\
       \texttt{sbm\_canonical\_mcmc} & \url{https://github.com/jg-you/sbm_canonical_mcmc} & 2019 \\
       \texttt{sbmr} & \url{https://github.com/tbilab/sbmr} & 2020 \\
       \texttt{sbmSDP} &\url{https://cran.r-project.org/web/packages/sbmSDP/} & 2015 \\
       \texttt{SBMSplitMerge} & \url{https://cran.r-project.org/web/packages/SBMSplitMerge/index.html} & 2020 \\
        \texttt{SparseBM} & \url{https://github.com/gfrisch/sparsebm} & 2021 \\\hline
    \end{tabular}
\end{table}
\begin{figure}
    \centering
    \includegraphics[width=1.0\linewidth]{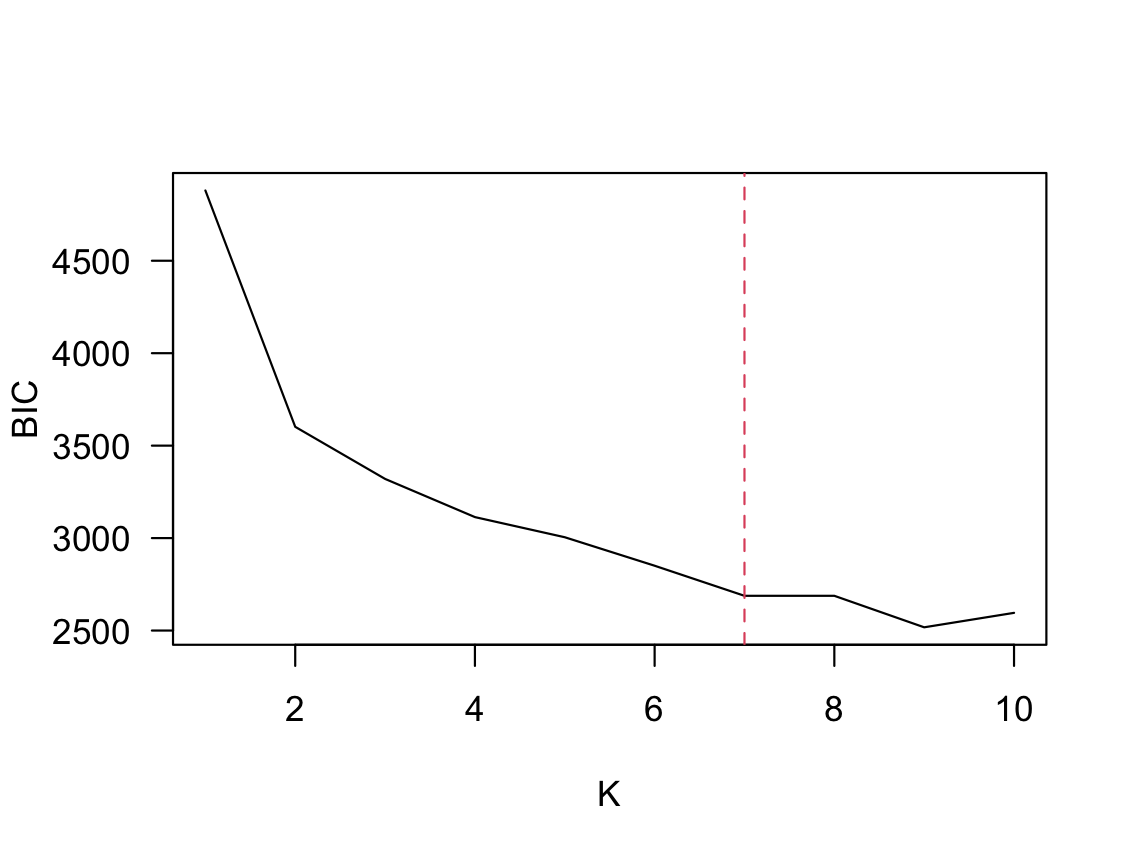}
    \caption{Elbow plot of BIC for the rUTI dataset by {Bayesian-SBM-MRF} ($f=1$). The elbow is at $K=7$ as indicated by the dashed red line.}
    \label{fig:bic_urinary}
\end{figure}
\begin{figure}
    \centering
    \includegraphics[width=1.0\linewidth]{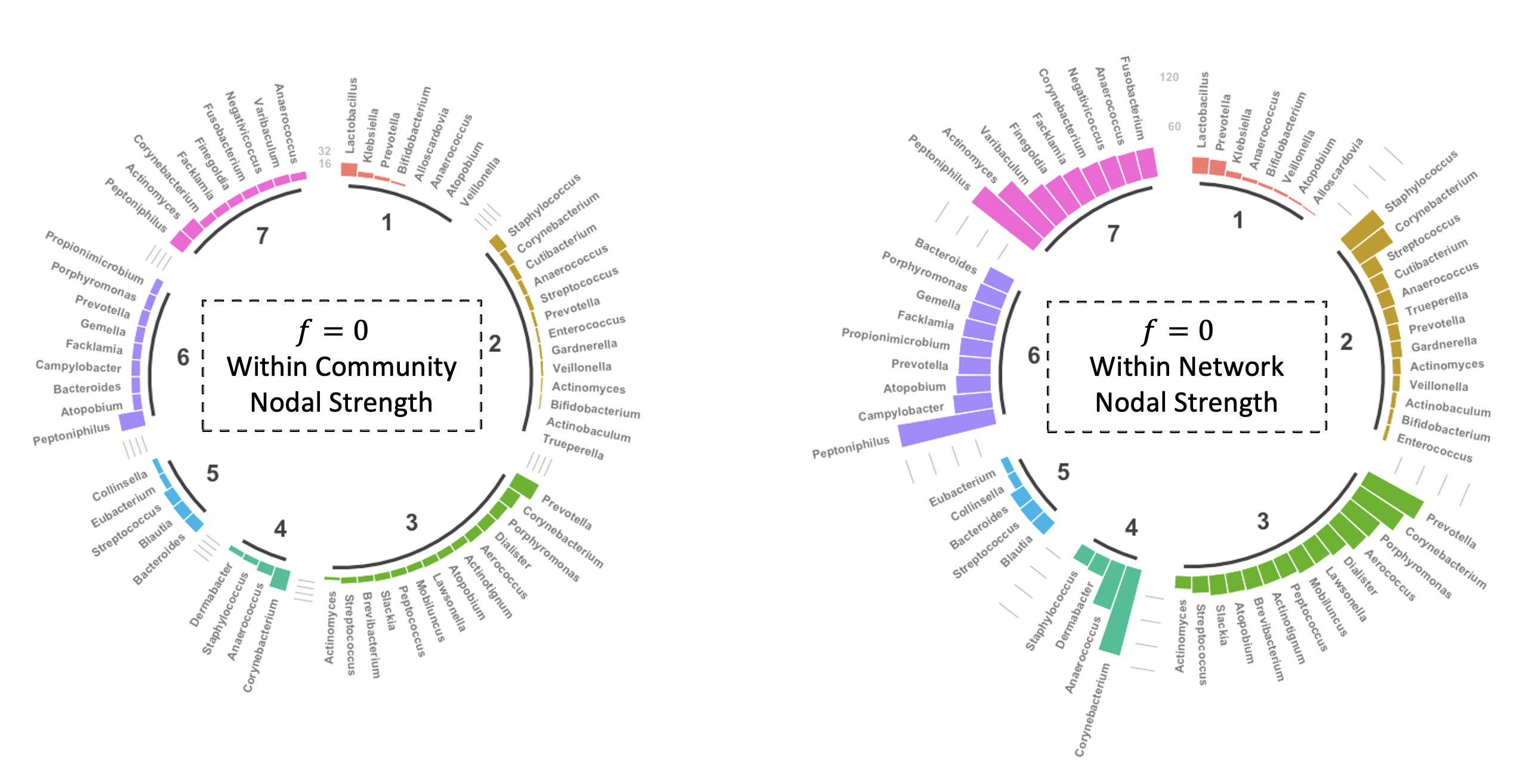}
    \includegraphics[width=1.0\linewidth]{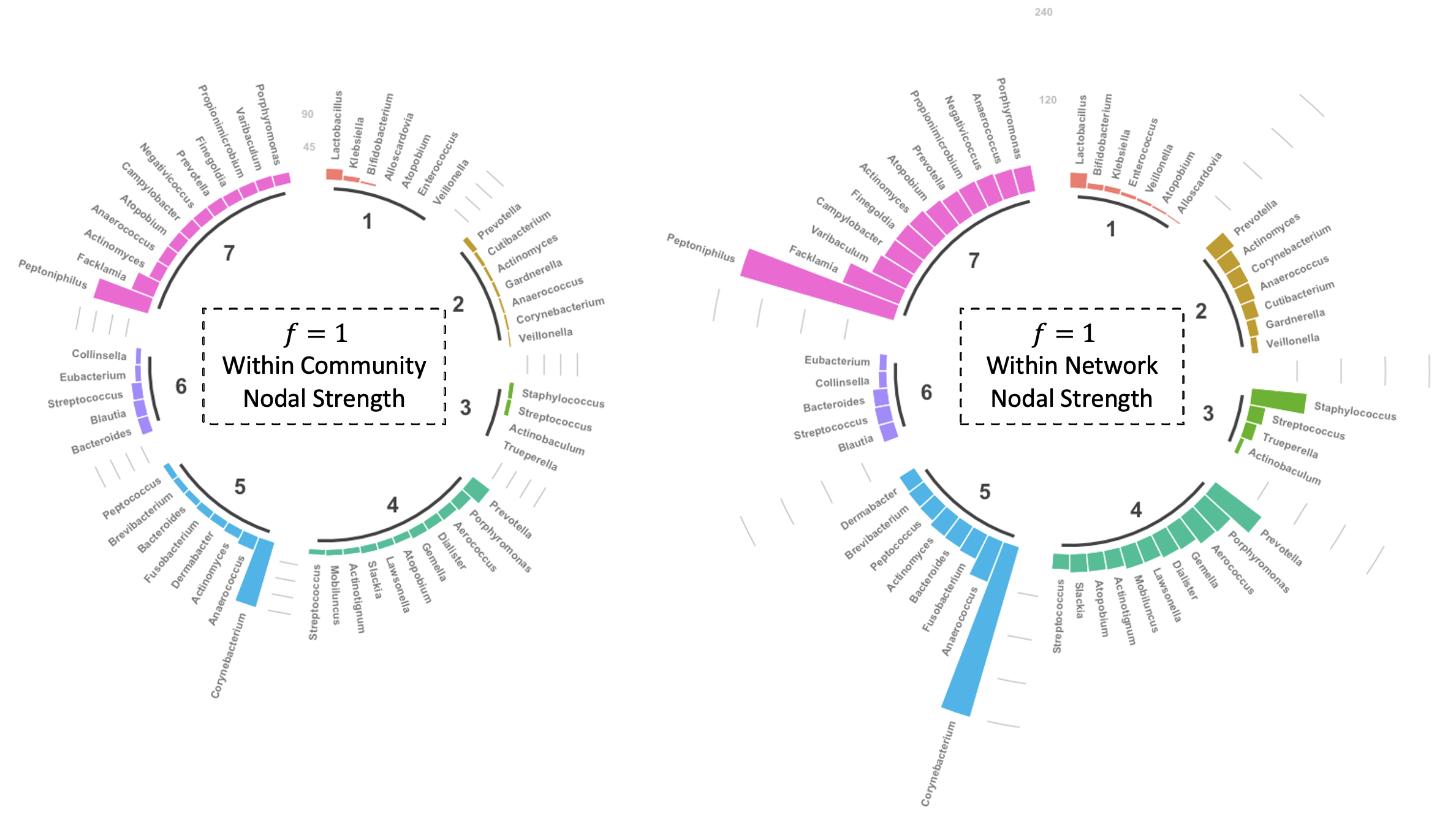}
    \caption{Circular bar plots of the nodal strength by genus within each community and within the entire network. The two top figures correspond to the standard Bayesian SBM ($f=0$) indicating that no taxonomic tree data have been incorporated into the model; the two bottom figures correspond to {Bayesian-SBM-MRF} ($f=1$) that incorporates taxonomic tree information.}
    \label{fig:both_sbm_f01}
\end{figure}
\begin{figure}
    \centering
    \includegraphics[width=1.0\linewidth]{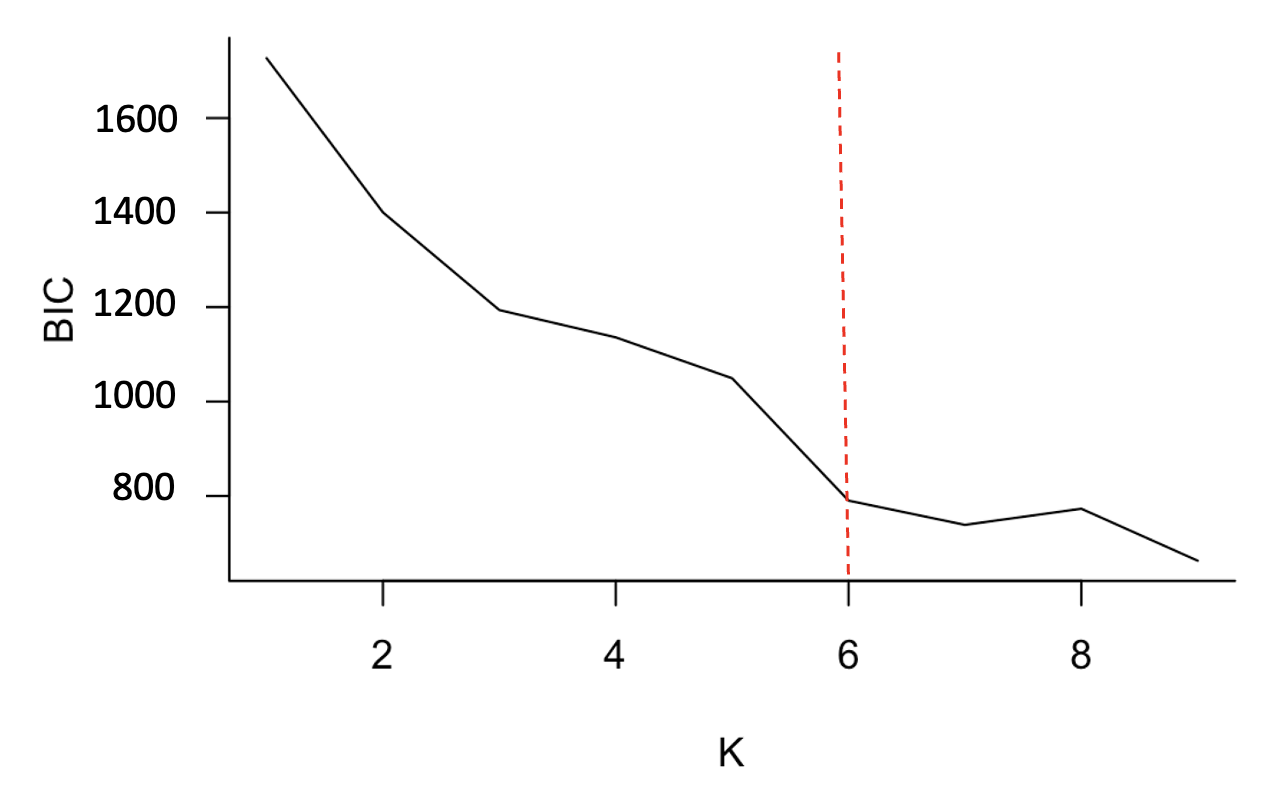}
    \caption{Elbow plot of BIC for the \textit{Les Mis\'erables} dataset by {Bayesian-SBM-MRF} ($f=1$). The elbow is at $K=6$ as indicated by the dashed red line.}
    \label{fig:bic_lesmis}
\end{figure}

\newpage
\bibliographystyle{plainnat}
%\bibliography{manuscript}

\end{document}